\documentclass[paper,11pt]{JHEP}
\usepackage[centertags]{amsmath}
\usepackage{amsfonts} \usepackage{amssymb} \usepackage{amsthm}
\allowdisplaybreaks[1]
\usepackage{graphicx}
\newcommand{\eq}[1]{Eq.~(\ref{#1})}

\newcommand{\ii}{\mathrm{i}}
\newcommand{\ee}{\mathrm{e}}

\newcommand{\diag}{\mathrm{diag}\,}

\newcommand{\cF}{{\mathcal{F}}}

\newcommand{\cN}{{\mathcal{N}}}

\newcommand{\cZ}{{\mathcal{Z}}}

\newcommand{\one}{{\rm 1\kern -.9mm l}}

\newcommand{\ft}[2]{{\textstyle\frac{#1}{#2}}}

\newdimen\tableauside\tableauside=1.0ex
\newdimen\tableaurule\tableaurule=0.4pt
\newdimen\tableaustep
\def\phantomhrule#1{\hbox{\vbox to0pt{\hrule height\tableaurule
width#1\vss}}}
\def\phantomvrule#1{\vbox{\hbox to0pt{\vrule width\tableaurule
height#1\hss}}}
\def\sqr{\vbox{%
  \phantomhrule\tableaustep
\hbox{\phantomvrule\tableaustep\kern\tableaustep\phantomvrule\tableaustep}%
  \hbox{\vbox{\phantomhrule\tableauside}\kern-\tableaurule}}}
\def\squares#1{\hbox{\count0=#1\noindent\loop\sqr
  \advance\count0 by-1 \ifnum\count0>0\repeat}}
\def\tableau#1{\vcenter{\offinterlineskip
  \tableaustep=\tableauside\advance\tableaustep by-\tableaurule
  \kern\normallineskip\hbox
    {\kern\normallineskip\vbox
      {\gettableau#1 0 }%
     \kern\normallineskip\kern\tableaurule}%
  \kern\normallineskip\kern\tableaurule}}
\def\gettableau#1 {\ifnum#1=0\let\next=\null\else
  \squares{#1}\let\next=\gettableau\fi\next}
\def\Xint#1{\mathchoice
   {\XXint\displaystyle\textstyle{#1}}%
   {\XXint\textstyle\scriptstyle{#1}}%
   {\XXint\scriptstyle\scriptscriptstyle{#1}}%
   {\XXint\scriptscriptstyle\scriptscriptstyle{#1}}%
   \!\int}
\def\XXint#1#2#3{{\setbox0=\hbox{$#1{#2#3}{\int}$}
     \vcenter{\hbox{$#2#3$}}\kern-.5\wd0}}
\def\ddashint{\Xint=}
\def\dashint{\Xint-}

\title{Modular anomaly equations in $\mathcal{N}=2^*$ theories and their large-$N$ limit 
}
\author{M. Bill\'o$^1$, M. Frau$^{1}$, F. Fucito$^{2}$, A. Lerda$^{1}$, J.F. Morales$^{2}$, R. Poghossian$^{3}$, D. Ricci Pacifici$^{4}$
\\
\vskip 0.2cm
$^1$ Universit\`a di Torino, Dipartimento di Fisica
\\ and I.N.F.N. - sezione di Torino, 
Via P. Giuria 1, I-10125 Torino, Italy\\
\vskip 0.2cm
$^2$ I.N.F.N - sezione di Roma 2\\
and Universit\`a di Roma Tor Vergata, Dipartimento di Fisica\\
Via della Ricerca Scientifica, I-00133 Roma, Italy\\
\vskip 0.2cm
$^3$Yerevan Physics Institute,
Alikhanian Br. 2, AM-0036 Yerevan, Armenia\\
\vskip 0.2cm
$^4$Universit\`a di Padova, Dipartimento di Fisica e Astronomia \\
and I.N.F.N - sezione di Padova, Via Marzolo 8, I-35131 Padova, Italy

\vspace{0.35cm}
\email{billo,frau,lerda@to.infn.it; fucito,morales@roma2.infn.it; poghos@yerphi.am; riccipacifici@pd.infn.it} 
}
\abstract{We propose a modular anomaly equation for the prepotential of the $\mathcal{N}=2^*$ super Yang-Mills theory 
on $\mathbb{R}^4$ with gauge group U($N$) in the presence of an $\Omega$-background. 
We then study the behaviour of the prepotential in a large-$N$ limit, in which $N$ goes
to infinity with the gauge coupling constant kept fixed. In this regime instantons are not suppressed.
We focus on two representative choices of gauge theory vacua, where the vacuum expectation values of the scalar fields are distributed either homogeneously or according to the Wigner semi-circle law. In both cases we derive an all-instanton exact formula 
for the prepotential. 
As an application, we show  that the gauge theory partition function on $\mathbb{S}^4$ at large $N$ localises around a Wigner 
distribution for the vacuum expectation values leading to a very simple expression in which the instanton contribution becomes independent of the coupling constant.   
}


\preprint{ROM2F/2014/05}

\begin{document}

\section{Introduction}
\label{secn:intro}
Holomorphic anomalies have made their first appearance in topological string theory in 
\cite{Bershadsky:1993cx,Bershadsky:1993ta} and since then they
have always received a lot of attention. In the context of field theories they were first 
studied for the $\mathcal{N}=2^*$ super Yang-Mills theory \cite{Minahan:1997if}, that is
a deformation of the maximally supersymmetric $\mathcal{N}=4$  gauge theory in which the adjoint 
hypermultiplet becomes massive. This theory can be regarded as an interpolation between the 
$\mathcal{N}=4$ model and the pure $\mathcal{N}=2$ super Yang-Mills theory 
to which it reduces by taking, respectively, the limit in which the hypermultiplet mass is sent to zero or decoupled. 
By giving a mass to the hypermultiplet, one breaks the  SL($2,\mathbb{Z}$) duality invariance of the 
original $\mathcal{N}=4$ theory, but a remnant of this symmetry manifests in the fact that the expansion coefficients 
of the $\mathcal{N}=2^*$ prepotential in the limit of small mass are almost modular forms of the bare gauge coupling. 
Quite remarkably, these coefficients satisfy a recursion relation \cite{Minahan:1997if} which translates into a partial differential 
equation for the prepotential itself. Such an equation, sometimes called modular anomaly equation, can be regarded as the gauge theory 
counterpart of the holomorphic anomaly equation satisfied by the topological string amplitudes, and can be used as a very efficient tool 
for a fast computation of the full prepotential order by order in the mass.
  
With the introduction of the so-called $\Omega$-background \cite{Moore:1998et,Nekrasov:2002qd,Nekrasov:2003rj} and 
its interpretation in terms of Neveu-Schwarz/Neveu-Schawarz or Ramond/Ramond field strengths of closed string theory 
\cite{Billo:2006jm,Hellerman:2011mv,Hellerman:2012zf,Orlando:2013yea}, the connection between the gauge theory prepotential
and the topological string amplitudes has become more clear, and recently there has been a renewed interest both in computing
such amplitudes \cite{Antoniadis:2010iq,Antoniadis:2011hq,Antoniadis:2013mna,Hohenegger:2013ala} and in adapting the 
previous works to this new scenario \cite{Huang:2011qx,KashaniPoor:2012wb,Billo:2013fi,Huang:2013eja,Billo:2013jba}. 
In particular, in \cite{Billo:2013fi,Billo:2013jba}
the $\cN=2^*$  theory with gauge group SU(2) has been studied in a generic $\Omega$-background characterized by two independent
parameters, $\epsilon_1$ and $\epsilon_2$, and its prepotential, including all its non-perturbative instanton corrections, 
has been computed in a small mass expansion. Moreover, it has been shown that 
the resulting quantum prepotential obeys a modular anomaly equation with a term proportional to $\epsilon_1\epsilon_2$, 
which generalizes the previously known equation of the underformed theory \cite{Minahan:1997if}.

The purpose of this paper is to extend these ideas to gauge theories with an arbitrary 
number $N$ of colors, and later study the large-$N$ limit. The large-$N$ limit of the $\cN=2^*$ theories with gauge group U($N$)
has been recently analised in \cite{Russo:2012ay,Russo:2013qaa,Russo:2013kea} using localization 
methods \cite{Pestun:2007rz}, and in \cite{Buchel:2013id,Buchel:2013fpa,Bobev:2013cja} 
using holography. In all these papers the $\cN=2^*$ theory is defined on the sphere $\mathbb{S}^4$, and 
the large-$N$ limit is taken by keeping fixed the 't~Hooft coupling $\lambda=Ng^2$.
In this limit, instanton configurations, which are weighted by 
the usual factor $q=\exp\left(-8\pi^2 /g^2\right)= \exp\left(-8\pi^2 N/\lambda\right)$, are exponentially suppressed 
and play no role%
\footnote{Actually, before reaching this conclusion, one should also to make sure that the integration over the instanton moduli space does not overcome the exponential suppression of the instanton weight, thus leading to an instanton-induced large-$N$ phase transition \cite{Gross:1994mr}. In \cite{Passerini:2011fe} it has been checked that this phenomenon does not occur
in the $\cN=2^*$ theory under consideration and that instantons remain exponentially suppressed in the 't~Hooft large-$N$ limit.}.

Here, instead we consider  the $\cN=2^*$ U($N$) theory on $\mathbb{R}^4$ and take 
the large-$N$ limit by keeping fixed the Yang-Mills coupling $g^2$. 
In this regime, sometimes called the very strongly coupled large-$N$ limit, the  weight $q$ remains \emph{finite} so that 
instantons cannot be discarded a priori \cite{Azeyanagi:2013fla} and their effects have to be consistently included in the picture.
Moreover, being on $\mathbb{R}^4$, the gauge theory should be supplemented by boundary conditions specifying how the fields behave at infinity. Focusing on the Coulomb branch characterised by the $N$ vacuum expectation values $a_i$'s for the scalar fields in the vector multiplet, we study how the prepotential behaves in the large-$N$ limit for different
choices of vacuum expectation values distributions. In particular we will consider a uniform distribution and the
Wigner semi-circle distribution, which both allow to perform explicit calculations and checks. 
In both cases we find that all instanton sectors contribute to the prepotential
in this large-$N$ limit and that the sum of all such contributions can be explicitly performed yielding a simple expression for 
the non-perturbative contribution to the prepotential at large $N$. It turns out that this contribution is subleading in $N$ with
respect to the classical and perturbative terms, and has some interesting properties.
    
Finally, we use these results to study the gauge theory prepotential on $\mathbb{S}^4$. In this case, the partition function 
can be written as  $Z_{\mathbb{S}^4}=\int da \big|Z_{\mathbb{R}^4}\big|^2 $ 
where the integral is over the vacuum expectation values $a_i$ of the scalar field in the
vector multiplet and  $Z_{\mathbb{R}^4}$ is related to the gauge theory prepotential in the $\Omega$ background 
on $\mathbb{R}^4$ via  $\cF=-\epsilon_1 \epsilon_2 \log Z_{\mathbb{R}^4}$. 
The large-$N$ limit of $Z_{\mathbb{S}^4}$ is evaluated using saddle-point techniques and the exact results obtained
for the prepotential on $\mathbb{R}^4$.  In particular we show that the integral over $a$ 
localises around a vacuum characterised by a Wigner distribution of the expectation values with a critical length leading
to a remarkably simple expression in which the instanton contribution to the prepotential becomes independent of the
coupling constant and hence negligible. 

The plan of the paper is as follows. In Section~\ref{secn:prep} we propose a holomorphic anomaly equation 
for the prepotential of the $\mathcal{N}=2^*$ theory 
with gauge group U($N$) in a generic $\Omega$-background. We display exact formulae for the first few terms
in the small mass expansion of the prepotential and check our results against those based on localization. 
In Sections~\ref{sec:largeN-sc} and \ref{sec:Wigner}, 
we analyse the large-$N$ limit for the two representative choices of gauge theory vacua 
where the scalar eigenvalues are distributed either homogeneously or  following the Wigner semi-circle law, and after resumming
over all instanton numbers we obtain the non-perturbative prepotential at large $N$.  
In Section~\ref{sec:concl} we present our conclusions and the results for the gauge theory partition function on $\mathbb{S}^4$.
Finally, in the Appendices we collect several technical details which are useful to reproduce the calculations presented in the main text.  

\section{The prepotential of the $\mathcal{N}=2^*$ theory}
\label{secn:prep}

The $\mathcal{N}=2^*$ theory is a massive deformation of the $\mathcal{N}=4$ super Yang-Mills theory describing the 
interactions of a $\mathcal{N}=2$ gauge vector multiplet with a massive $\mathcal{N}=2$ hypermultiplet in the adjoint representation.
In the following we consider the $\mathcal{N}=2^*$ theory with gauge group $\mathrm{U}(N)$ and mass $m$ in 
an $\Omega$-background \cite{Nekrasov:2002qd,Nekrasov:2003rj} parameterized by $\epsilon_1$ and $\epsilon_2$ which, for later
convenience, we combine in
\begin{equation}
\epsilon \equiv    \epsilon_1+\epsilon_2 ~~~\mbox{and}~~~h\equiv\sqrt{\epsilon_1\epsilon_2}~.
\label{epsilonh}
\end{equation}
In the vacuum where the scalar field $\Phi$ of the vector multiplet has expectation value
\begin{equation}
\langle\Phi\rangle = \diag\big(a_1,\ldots,a_N\big)~,
\label{vev}
\end{equation}
the prepotential $\cF$ can be written as 
\begin{equation}
\cF = \cF^{\mathrm{class}}+ \cF^{\mathrm{quant}}
\label{prepot}
\end{equation}
where the classical term is simply
\begin{equation}
\cF^{\mathrm{class}} = \pi\ii \tau\sum_{i=1}^Na_i^2~,
\label{Fclass}
\end{equation}
with
\begin{equation}
\label{tau}
\tau=\frac{\theta}{2\pi}+\ii\frac{4\pi}{g^2}
\end{equation}
being the (complexified) gauge coupling constant.

The quantum prepotential is the sum of a perturbative 1-loop term and a non-perturba- tive part due to instantons:
\begin{equation}
\cF^{\mathrm{quant}}=\cF^{\mathrm{1-loop}}+\cF^{\mathrm{inst}}~.
\label{prepotquantum}
\end{equation}

\subsection{The one-loop prepotential}

The 1-loop piece for the $\mathcal{N}=2^*$ theory is given by \cite{Nekrasov:2003rj,Huang:2011qx,Billo:2013fi}
\begin{equation}
\cF^{\mathrm{1-loop}}= h^2 \sum_{i\not= j}\Big[\log\Gamma_2(a_{ij})-\log\Gamma_2(a_{ij}+m+\epsilon)\Big]
\label{F1loop}
\end{equation}
where $a_{ij}=a_i-a_j$ and $\Gamma_2$ is the Barnes double $\Gamma$-function (see Appendix \ref{appd} for our notations
and conventions). Expanding for small values of $m$, $h$ and $\epsilon$, one gets
\begin{equation}
\cF^{\mathrm{1-loop}}= \sum_{n=1}^\infty f_n^{\mathrm{1-loop}}
\label{F1loopexp}
\end{equation}
where $f_n^{\mathrm{1-loop}}$ is a homogeneous polynomial of order $2n$ in $(m,h,\epsilon)$.  The first few terms are
\begin{subequations}
\begin{align}
f_1^{\mathrm{1-loop}}&=\frac{M^2}{4}\sum_{i\not=j} \log\left( \frac{a_{ij}}{\Lambda}\right)^2~,\label{f11loop}\\
f_2^{\mathrm{1-loop}}&=-\frac{M^2(M^2+h^2)}{24}\,C_2~,\label{f21loop}\\
f_3^{\mathrm{1-loop}}&=-\frac{M^2(M^2+h^2)\big(2M^2-\epsilon^2+3 h^2\big)}{240}\,C_4~.\label{f31loop}
\end{align}
\label{fn1loop}
\end{subequations}
Here we have defined
\begin{equation}
M^2=m^2-\frac{\epsilon^2}{4}~,
\label{M2}
\end{equation}
and introduced the sums
\begin{equation}
C_n=\sum_{i\not=j}\frac{1}{(a_{ij})^n}~,
\label{Cn}
\end{equation}
and an arbitrary scale $\Lambda$ in the logarthmic term.
The coefficients $f_n^{\mathrm{1-loop}}$'s for higher $n$
have more complicated expressions but their dependence on the vacuum expectation values
$a_i$'s is entirely through the sums $C_n$. 
For a vanishing $\epsilon$-background, (\ref{F1loop}) reduces to%
\footnote{Notice that all terms with odd powers of $m$ and hence of $a_i$ vanish when summed over $i\not=j$.} 
\begin{eqnarray}
\cF^{\mathrm{1-loop}}&=& \frac{1}{4}\sum_{i\not= j}\Big[
-a_{ij}^2\log\Big(\frac{a_{ij}}{\Lambda }\Big)^2 + (a_{ij}+m)^2\log\Big(\frac{a_{ij}+m}{\Lambda}\Big)^2
+ 3 a_{ij}^2 - 3 (a_{ij}+m)^2
\Big]\nonumber\\
&=&\frac{m^2}{4} \sum_{i\not=j} \log\Big(\frac{a_{ij}}{\Lambda} \Big)^2 - 
\sum_{n=2}^\infty \frac{m^{2n}}{4n(n-1)(2n-1)}\, C_{2n-2}~,
\label{F1lnoe}
\end{eqnarray}
so that, in this case, we simply have
\begin{equation}
\label{f1lnoe}
f_n^{\mathrm{1-loop}} = - \frac{m^{2n}}{4n(n-1)(2n-1)}\, C_{2n-2}
\end{equation}
for any $n>1$, while $f_1^{\mathrm{1-loop}}$ is given by (\ref{f11loop}) with $M^2$ replaced by $m^2$.

\subsection{The instanton prepotential}

The instanton part $\cF^{\mathrm{inst}}$ can be computed order by order in the instanton counting parameter 
\begin{equation}
q = \ee^{2\pi\ii\tau}
\label{q}
\end{equation}
using localization methods \cite{Moore:1998et,Nekrasov:2002qd,Nekrasov:2003rj}. 
In this framework, the instanton prepotential is viewed  as the free energy 
\begin{equation}
 \cF^{\mathrm{inst}}= - \epsilon_1 \epsilon_2  \log  Z^{\mathrm{inst}}=  \sum_{k=1}^\infty q^{k}\,\cF^{(k)}
\label{Finst0}
\end{equation}
of a statistical system with a (gran canonical) partition function $Z^{\mathrm{inst}}=1+\sum_{k=1}^\infty q^{k}\, Z_k$,
counting the number of excitations in the instanton moduli space.
 
Referring for example to \cite{Bruzzo:2002xf,Fucito:2011pn,Billo:2012st} for details, the $k$-instanton partition function
of the $\cN=2^*$ theory is given by
\begin{equation}
 \begin{aligned}
Z_k &= (-1)^{k} \oint \prod_I^k \frac{d\chi_I}{2 \pi \ii }\, 
\prod_{I,J}^k  \left[ 
\frac{ (\chi_{IJ}+\delta_{IJ}) (\chi_{IJ}+\epsilon)}{(\chi_{IJ}+\epsilon_1)(\chi_{IJ}+\epsilon_2)}\,
\frac{( \chi_{IJ}+\epsilon_1+m )( \chi_{IJ}+\epsilon_2+m )}{(\chi_{IJ}+m)(\chi_{IJ}+\epsilon+m )} \right] \\
&~~~~~~~~~~\times  \prod_{I=1}^k \prod_{i=1}^N
\left[  \frac{ -(\chi_I-a_i)^2+\ft{(\epsilon+2 m)^2}{4}}{-(\chi_I-a_i)^2+\ft{\epsilon^2}{4}}  \right] 
\end{aligned}
\label{zk}
\end{equation}
where $\chi_{IJ}=\chi_I-\chi_J$, with $\chi_I$ being the instanton moduli that, in the U($k$) theory, are the analogue of the 
scalar vacuum expectation values $a_i$. At $k=1$ the instanton prepotential is 
\begin{equation}
\begin{aligned}
\cF^{(1)}&= -\epsilon_1\epsilon_2 \log Z_1\\
&=-\big(m+\frac{\epsilon}{2}-\epsilon_1\big)\big(m+\frac{\epsilon}{2}-\epsilon_2\big)
\sum_{i=1}^N\prod_{j\not=i}\frac{(a_{ij}+m+\ft{\epsilon}{2})(a_{ji}+m-\ft{\epsilon}{2})}{a_{ij}(a_{ji}-\epsilon)}~.
\end{aligned}
\label{F1inst} 
\end{equation}
The localization techniques allow to push the calculation to higher instanton numbers without major 
problems, but the resulting explicit expressions for the $\cF^{(k)}$'s
quickly become rather cumbersome. However, expanding 
for small values of $m$, $h$ and $\epsilon$  as in (\ref{F1loopexp}),
it is possible to write the homogeneous polynomials $f_n^{\mathrm{inst}}$ in a reasonably compact way. 
For example, 
up to three instantons we find
\begin{subequations}
\label{fninst}
\begin{align}
f_1^{\mathrm{inst}}\!&=\! -N(M^2+h^2)\Big(q+\frac{3}{2}\,q^2+\frac{4}{3}\,q^3+\ldots\Big)~,\label{f1inst}\\
f_2^{\mathrm{inst}}\!&=\!M^2(M^2+h^2)\big(q+3\,q^2+4\,q^3+\ldots\big)C_2~,\label{f2inst}\\
f_3^{\mathrm{inst}}\!&=\!M^2(M^2+h^2)\Big[ \epsilon^2\, q-3(2M^2-3\epsilon^2+3h^2)q^2-4(8M^2-7\epsilon^2+12h^2)
q^3+\ldots\Big]C_4\nonumber\\&
~~~~-\frac{1}{2}M^4(M^2+h^2)\big(q^2+6\,q^4+12\,q^6+\ldots\big)C_{22}~.\label{f3inst}
\end{align}
\end{subequations}
Of course there are no obstructions in obtaining the instanton expansions of the higher $f_n^{\mathrm{inst}}$'s,
but we do not write them here since their explicit expressions will not be needed in the following.
A few comments are in order. First, we notice that there are no non-perturbative corrections
to the logarithmic term (\ref{f11loop}), as expected, and that $f_1^{\mathrm{inst}}$ is 
independent of the vacuum expectation values $a_i$'s and that its
$k$-th instanton coefficient is $\sigma_1(k)/k$ where $\sigma_1(k)$ is the sum of the divisors of $k$. 
Second, we see that new structures, which were not present in the 1-loop results (\ref{fn1loop}), start appearing in 
the non-perturbative sector; for example $f_3^{\mathrm{inst}}$ contains the triple sum
\begin{equation}
C_{nm}=\sum_{i\not=j\not=k\not=i} \frac{1}{(a_{ij})^n\,(a_{ik})^m}
\label{Cnm}
\end{equation}
with $n=m=2$. More and more structures appear in the higher $f_n^{\mathrm{inst}}$'s. For instance, the
exact 1-instanton contribution (\ref{F1inst}) in the undeformed case ($\epsilon_1,\epsilon_2\to 0$)
can be written as\footnote{Here we use the shorthand $0!=1$. }
\begin{equation}
\label{oneinst1}
\cF^{(1)}= - m^2 \sum_{i=1}^N\,\prod_{j\not=i}\Big(1 - \frac{m^2}{a_{ij}^2}\Big)= \sum_{n=1}^N \frac{(-1)^n\,m^{2n}}{(n-1)!}~
C_{\underbrace{\mbox{\scriptsize{22\ldots2}}}_{\mbox{\scriptsize{$n-1$}}}}~,
\end{equation}
where 
\begin{equation}
C_{\underbrace{\mbox{\scriptsize{22\ldots2}}}_{\mbox{\scriptsize{$n-1$}}}}
~= \sum_{[i_1,i_2, \ldots,i_n]} \frac{1}{(a_{i_1 i_2})^2\,(a_{i_1 i_3})^2 \ldots (a_{i_1 i_n})^2}~,
\label{C222n}
\end{equation}
and the symbol $[i_1,i_2, \ldots,i_n]$ denotes a sum over $n$ non-coinciding positive integers (see Appendix \ref{appa} for details).    

 \subsection{The exact prepotential}

The sum of the perturbative and non-perturbative contributions (\ref{fn1loop}) and (\ref{fninst}), which we denote simply as $f_n$, 
can be recast in a more compact and suggestive way. In fact, building on previous results obtained both in the U($N$) undeformed 
$\cN=2^*$ theory  \cite{Minahan:1997if}
and in the SU(2) deformed one \cite{Huang:2011qx,KashaniPoor:2012wb,Billo:2013fi,Huang:2013eja,Billo:2013jba}, 
we expect that $f_n$ (with $n>1$) are (quasi) modular forms of 
weight $(2n-2)$ which can be expressed as homogeneous polynomials in the Eisenstein 
series $E_{2n}$ (see Appendix \ref{appb} for our conventions). Indeed we have
\begin{subequations}
\begin{align}
f_1&= \frac{M^2}{4}\sum_{i\not=j} \log\left( \frac{a_{ij}}{\Lambda}\right)^2+N\,(M^2+h^2)\log \widehat \eta(q)  ~,\label{f1t}\\
f_2&=-\frac{M^2(M^2+h^2)}{24}\,E_2 \,C_2~,\label{f2t}\\
f_3&=-\frac{M^2(M^2+h^2)}{288}\Big[(2M^2+3h^2)E_2^2+\frac{1}{5}(2M^2-6\epsilon^2+3h^2)E_4\Big] C_4\\
&~~~~+\frac{M^4(M^2+h^2)}{576}\big(E_2^2-E_4\big) C_{22}~.\label{3t}
\end{align}
\label{f23}
\end{subequations}
In (\ref{f1t}) we have used the identity
\begin{equation}
\sum_{k=1}^\infty \frac{\sigma_1(k)}{k}\,q^{k} = -\log \widehat \eta(q) 
\label{sigmaeta}
\end{equation}
to rewrite the instanton contribution of $f_1$ in terms of the Dedekind $\eta$-function and defined
$\widehat{\eta}(q)=q^{-{1/24}} \eta (q)$.
Expanding the modular functions for small values of $q$, one can easily check that both the 
perturbative part (\ref{fn1loop}) and the first few instanton corrections 
(\ref{fninst}) are correctly reproduced by (\ref{f23}).

\subsection{Recursion relation}
The explicit expressions (\ref{f23}) allow us to verify the following non-linear relation
\begin{equation}
\partial_{E_2} f_n=-\frac{1}{24}\sum_{m=1}^{n-1} \vec{\nabla} f_m\cdot \vec{\nabla} f_{n-m}+\frac{h^2}{24}\,\Delta f_{n-1}
\label{recursion}
\end{equation}  
for $n=1,2,3$, where $\vec{\nabla}=\big(\partial_{a_1},\partial_{a_2},\ldots,\partial_{a_N}\big)$ 
and $\Delta=\sum_i \partial_{a_i}^2$. To prove this relation one has to use 
various identities, like for example 
\begin{equation}
C_{11}=0~,~~~~C_{31}=-\frac{1}{4}\,C_{22}~,
\label{identities}
\end{equation}
which can be easily checked as discussed in Appendix \ref{appa}. 

Eq.~(\ref{recursion}) is a generalization to U($N$) of the recursion relation found in \cite{Billo:2013fi,Billo:2013jba} 
for the $\epsilon$-deformed SU(2) theory,  and a generalization in presence of the $\Omega$-background of the relation 
found in \cite{Minahan:1997if} for the undeformed U($N$) theory\,\footnote{The results reported in \cite{Minahan:1997if} contain two minor misprints: one in the recursion equation for the $f_n$'s and one regarding a factor of 1/4 in the last term of $f_4$.}. As discussed in those references, the recursion relation obeyed by the prepotential coefficients
can be regarded as a direct consequence of the modular anomaly equation
which, in turn, encodes the same information as the holomorphic anomaly equation of the topological string amplitudes
\cite{Bershadsky:1993cx,Bershadsky:1993ta} and its generalizations \cite{Huang:2011qx,KashaniPoor:2012wb,Billo:2013fi,Huang:2013eja,Billo:2013jba}. Indeed, the holomorphic anomaly equation
implies an anomalous modular behavior of the prepotential coefficients with respect to $\tau$ 
which can only occur through the Eisenstein series $E_2$. 

We can therefore view (\ref{recursion}) as
a distinctive property of the $\cN=2^*$ U$(N)$ theory which is valid for any $n$, and use it to derive recursively 
the various polynomials $f_n$ for $n>3$. To this purpose one has to choose a basis for the sum structures that appear
for high values of $n$ and fix the integration constants. We do this by choosing to write everything only in terms of sums 
involving even inverse powers of $a_{ij}$, $a_{ik}$, $\ldots$, which
is always possible due to the existence of algebraic identities like those displayed 
in (\ref{identities}) or in Appendix \ref{appa}, and fix 
the $E_2$ independent terms by comparing with the perturbative and the first-few instanton expressions. Proceeding
in this way, we have computed $f_n$ up to $n=6$. The result for $f_4$ is given in Appendix \ref{appc} (see (\ref{f4}))
including its full dependence on the $\Omega$-background parameters; in the same Appendix (see (\ref{f5}) and (\ref{f6})),  
we also write $f_5$ and $f_6$ for the undeformed $\cN=2^*$ U($N)$ theory.

Finally, we observe that the recursion relation (\ref{recursion}) implies the following differential equation for the
quantum prepotential:
\begin{equation}
\partial_{E_2} \cF^{\mathrm{quant}}
=-\frac{1}{24}\vec{\nabla} \cF^{\mathrm{quant}}\cdot \vec{\nabla}\cF^{\mathrm{quant}}+\frac{h^2}{24}\,\Delta \cF^{\mathrm{quant}}
\label{recursion1}
\end{equation}
which is the homogeneous Kardar-Parisi-Zhang equation in $N$ space dimensions.
By introducing the quantum partition function
\begin{equation}
\cZ^{\mathrm{quant}}= \exp\Big({-\frac{\cF^{\mathrm{quant}}}{h^2}}\Big)~,
\label{Z}
\end{equation}
we can map the above non-linear equation into a (parabolic) linear one, namely
\begin{equation}
\partial_{E_2} \cZ^{\mathrm{quant}}
-\frac{h^2}{24}\,\Delta \cZ^{\mathrm{quant}}=0~,
\end{equation}
which is the heat equation in $N$ space dimensions.

\section{Large-$N$ limit with a uniform distribution}
\label{sec:largeN-sc}

We now discuss the large-$N$ limit of the $\cN=2^*$ theory, focusing in particular on its non-perturbative sector. The first regime we consider is a naive semi-classical configuration \cite{Douglas:1995nw} in which every charged multiplet has a mass of 
order $N^0$ or greater.
This is possible if we choose the eigenvalues $a_i$ in such a way that the minimum difference among them is $O(N^0)$.
A representative choice in this regime is  
\begin{equation}
a_i=v \Big(i-\frac{N+1}{2}\Big)\quad\mbox{for}~i=1,\ldots,N~,
\label{vevsemiclassical}
\end{equation}
where $v$ is a constant scale carrying the physical dimension of a mass. Such a distribution 
has been considered several times in the literature and its properties have been discussed in various contexts, see for 
example \cite{Douglas:1995nw,Ferrari:2001mg}.

If $v$ is constant, the highest eigenvalue is $O(N)$ and the others are equally spaced in an interval $[-\mu,\mu]$ where 
$\mu$ is  $O(N)$. {From} (\ref{vevsemiclassical}) we see that $\sum_i a_i=0$ so that this distribution actually
applies to SU($N$) gauge groups. Moreover, it is easy to check that the classical prepotential
(\ref{Fclass}) grows as $N^3$ when it is evaluated with (\ref{vevsemiclassical}). 

It is also interesting to observe that among all possible distributions of $a_i$ in a range of size
$O(N)$, the uniform distribution (\ref{vevsemiclassical}) seems to be the one which minimizes the various sums $C_n$, $C_{nm}$ and
so on, which we introduced in the previous section. We have checked this property numerically in several examples. 
We therefore think that it is worth investigating the large-$N$ behavior of the $\cN=2^*$ SU($N$) theory
with the uniform distribution (\ref{vevsemiclassical}) which, given its simplicity, will allow us also to obtain quite explicit results.

\subsection{The pure SU$(N)$ theory}
\label{subsec:pure}
We begin by considering the pure $\cN=2$ super Yang-Mills theory with gauge group SU($N$), whose large-$N$ behaviour was
studied long ago in \cite{Douglas:1995nw}. In the pure theory a scale $\Lambda$ is dynamically generated and the $k$-th instanton sector contributes at order $\Lambda^{2Nk}$. 
One then typically studies 1-instanton effects to understand if, and when, they can survive in the large-$N$ limit
and signal the breakdown of the semi-classical perturbative regime dominated the 1-loop term.

Considering the uniform distribution (\ref{vevsemiclassical}) and exploiting the explicit form of the Seiberg-Witten curve for the 
pure theory, the 1-instanton contributions to the prepotential at large $N$ were estimated in 
\cite{Douglas:1995nw} to be of order
\begin{equation}
\label{f1largeNfin}
q\,\cF^{(1)}\Big |_{\mathrm{pure}\,\mathrm{SU}(N)} \sim   \Big(\frac{\Lambda}{Nv}\Big)^{2N}~,
\end{equation}
This shows that instanton contributions in this regime are strongly suppressed, not 
just because of their dependence $\Lambda^{2N}$ on the dynamical scale $\Lambda$ but also by an extra $N^{-2N}$
dependence. Thus instantons do not contribute even at strong coupling ( $\Lambda \sim 1$ ) unless we take $v $ of order $\Lambda/N$.  For this choice the range of values of the $a_i$'s in (\ref{vevsemiclassical}) stays finite inside a region of length $\Lambda$,
and the differences $a_{ij}$ (related to the $W$-boson masses) vanish at large $N$.

The result (\ref{f1largeNfin}) can be retrieved from the explicit form of the $\cN =2^*$ prepotential given in the previous
section by taking a \emph{decoupling limit} in which the mass of the adjoint hypermultiplet diverges and 
the gauge coupling vanishes keeping finite the product $m^{2N}\, q$, 
 namely
\begin{equation}
\label{decoupling1}
m\to\infty~,~~~
q\to 0~,~~~\mathrm{with}~~
\Lambda^{2N} \equiv (-1)^N\, m^{2N}\, q~~\mathrm{finite}~.
\end{equation}
This check makes us confident in deriving later the large-$N$ limit of instanton contributions in the $\cN=2^*$ theory, 
where a derivation based on the Seiberg-Witten curve is not available. 

In the decoupling limit (\ref{decoupling1}) only the terms with the highest mass power in (\ref{oneinst1}) survive,
and thus the 1-instanton contribution to the prepotential of the pure SU($N$) theory is
\begin{equation}
\label{f1largeN}
q \,\cF^{(1)}\Big |_{\mathrm{pure}\,\mathrm{SU}(N)} ~\sim ~ \frac{\Lambda^{2N}}{(N-1)!} ~
C_{\underbrace{\mbox{\scriptsize{22\ldots2}}}_{\mbox{\scriptsize{$N-1$}}}}~.
\end{equation}
One can check that this expression exactly agrees with the explicit 1-instanton prepotential obtained 
from the Seiberg-Witten curve (see for example \cite{D'Hoker:1996nv}).

Exploiting the results of Appendix \ref{subappa:unif} we can now evaluate the behavior of the above expression in
the large-$N$ limit with the uniform distribution (\ref{vevsemiclassical}). In particular, from (\ref{c222Na}) we 
see that
\begin{equation}
\label{C222N}
C_{\underbrace{\mbox{\scriptsize{22\ldots2}}}_{\mbox{\scriptsize{$N-1$}}}}~\,\underset{N\to \infty} {\sim} ~
\frac{2N (N-1)!}{(2N)!}\,\Big(\frac{2\pi}{v}\Big)^{2N-2}~.
\end{equation}
Plugging this result into (\ref{f1largeN}), using the Stirling approximation and absorbing all numerical constants into $\Lambda$,
the estimate ( \ref{f1largeNfin}) is reproduced.  In the following we generalize this analysis to the $\cN=2^*$ theory and 
evaluate the large-$N$ limit at all-instanton orders.   
 
\subsection{The $\cN=2^*$ theory}
\label{subsec:n2*unif}
If we do not take the decoupling limit (\ref{decoupling1}) and consider the full $\cN=2^*$ theory, 
then all mass structures contribute to the prepotential and the complete expression of the latter 
is necessary to find the large-$N$ behavior of the theory. Recalling that the prepotential
coefficients $f_n$'s depend on the expectation values $a_i$ through the sums introduced in (\ref{sums}), we can use the
results of Appendix \ref{subappa:unif} to estimate how the $\cN=2^*$ prepotential behaves in the large-$N$ limit with
the uniform distribution (\ref{vevsemiclassical}). Basically what we have to do is to plug (\ref{sumcnN}) and (\ref{sumN})
into (\ref{f23}) (where now we take $h=\epsilon=0$) and into (\ref{f4}) -- (\ref{f6}) up to $n=6$. This amounts, for instance,
to replace the various sums according to
\begin{equation}
 \big(C_2, C_4, C_{22}, \ldots\big) ~\, \underset{N\to \infty} {\sim}  ~  N \, 
 \left(  \frac{2\zeta(2)}{v^2}, \frac{2\zeta(4)}{v^4},  \frac{4\zeta(2)^2-2 \zeta(4)}{v^4}, \ldots \right)~.
 \end{equation}
 Doing this, we find that numerous remarkable cancellations take place: 
indeed, in $f_3,\ldots,f_6$ all terms involving $E_2$ cancel and we remain with the following very simple result:
\begin{equation}
\label{f36lim}
\begin{aligned}
f_2& \,\underset{N\to \infty} {\sim}   -N\, \frac{m^4}{v^2}\,\frac{\zeta(2)}{12}\,E_2~, \\
f_3 &  \,\underset{N\to \infty} {\sim} -N\, \frac{m^6}{v^4}\,\frac{\zeta(4)}{60}\,E_4~,\\
f_4 &  \,\underset{N\to \infty} {\sim} -N\,\frac{m^8}{v^6}\,\frac{\zeta(6)}{168}\,E_6~,\\
f_5 &  \,\underset{N\to \infty} {\sim} 
-N\,\frac{m^{10}}{v^8}\,\frac{\zeta(8)}{360}\, E_8~,\\
f_6 & \,\underset{N\to \infty} {\sim} 
 -N\,\frac{m^{12}}{v^{10}}\,\frac{\zeta(10)}{660}\,E_{10} ~.
\end{aligned}
\end{equation}
It is very tempting to generalize these formulas to any $n>1$ by writing
\begin{equation}
\label{fnlim}
f_n  \,\underset{N\to \infty} {\sim} -N\,\frac{m^{2n}}{v^{2n-2}}
\,\frac{\zeta(2n-2)}{2n(n-1)(2n-1)}\, E_{2n-2}~.
\end{equation}
This extrapolation is actually not at all obvious, even if we assume that for $n>2$ all contributions containing 
$E_2$ factors cancel as they do up to $n=6$, and that the result is a modular form of weight $2n-2$. 
Indeed, for $n=3,\ldots, 6$ (and for $n=8$) 
the space of modular forms of weight $2n-2$ is one-dimensional, and if we assume no $E_2$ dependence except in $f_2$, then
the $f_n$'s \emph{have} to be proportional to $E_{2n-2}$ as we have found in (\ref{f36lim}).
However, for higher values of $n$, the corresponding space of modular forms has dimension greater than one; 
for instance, for $n=7$, $E_4^3$ and $E_6^2$ represent two independent modular forms of weigth 12 and in principle any linear
combination of them could appear in $f_7$. Thus, in general it is not at all guaranteed that the result should be simply proportional to 
the particular modular form $E_{2n-2}$. 
Nonetheless, we will present strong evidence in favour of  the conjecture (\ref{fnlim}) by showing a perfect match 
 of its predictions at the perturbative and 1-instanton level with the large-$N$ limit of the 
exact perturbative and 1-instanton results presented in 
Section \ref{secn:prep}. This check is actually conclusive only for the values of $n$ such that the space of modular forms of
weight $2n-2$ is two-dimensional; yet, also in the general case, the agreement that we find at this level is already very far from 
trivial.

Let us now give some details. Up to order $q$, {\it i.e.} up to 1-instanton, 
the Eisenstein series $E_{2n-2}$ are given by (see Appendix \ref{appb} and in particular (\ref{eis2nq1}))
\begin{equation}
\label{eisq2}
E_{2n-2} = 1 + \frac{2}{\zeta(3-2n)}\,q+ \ldots
\end{equation}  
for any $n>1$. Inserting this into (\ref{fnlim}), we easily see that our conjecture predicts a perturbative part for $f_n$ given by
\begin{equation}
\label{fn1loop1}
f_n^{\mathrm{1-loop}} \,\underset{N\to \infty} {\sim} -N\,\frac{m^{2n}}{v^{2n-2}}
\,\frac{\zeta(2n-2)}{2n(n-1)(2n-1)}~,
\end{equation}
and a 1-instanton part that can be expressed as 
\begin{equation}
\label{fn1}
f_n^{(1)}\,\underset{N\to \infty} {\sim}
  N\,\frac{m^{2n}}{v^{2n-2}}\,\frac{ \zeta(2n-2)}{(n+1)(2n+1)\zeta(1-2n)     }          ~,
\end{equation}

On the other hand, $f_n^{\mathrm{1-loop}}$ and $f_n^{(1)}$ can be extracted from
the \emph{exact} expressions of the perturbative and 1-instanton prepotential presented in Section \ref{secn:prep}.
The exact 1-loop contribution is given in (\ref{f1lnoe}) and, after taking into account the uniform large-$N$ behavior (\ref{sumcnN})
of the sums $C_n$, we easily see that it precisely agrees with the prediction (\ref{fn1loop1}) 
that follows from the ansatz (\ref{fnlim}).

The 1-instanton coefficients $f_n^{(1)}$ can be extracted from (\ref{oneinst1}) after estimating the sum 
$C_{22\ldots 2}$ in the large-$N$ limit. This can be done using (\ref{c222N}) and the coefficients $\alpha_n$
given in table (\ref{alphatable}). Then, for any $n>1$ we find
\begin{equation}
\label{fn1exact}
f_n^{(1)} = (-1)^nm^{2n}\, \frac{1}{(n-1)!}
\,C_{\underbrace{\mbox{\scriptsize{22\ldots2}}}_{\mbox{\scriptsize{$n-1$}}}}\,~\underset{N\to \infty} {\sim}~
(-1)^n N\,\frac{m^{2n}}{v^{2n-2}}\,\frac{\alpha_{n-1}\,\zeta(2n-2)}{(n-1)!}~.
\end{equation}
Comparing the result with (\ref{fn1exact}) with (\ref{fn1}) we  find perfect agreement %
\footnote{The computation of the $\alpha_n$'s can be straightforwardly extended to arbitrary $n$, finding all the times
agreement with (\ref{fn1}); it is therefore very natural to conjecture that the two formulas coincide and that 
$\alpha_n$ is given by (\ref{alphares}) for all $n$'s.}. 
Thus we can conclude that our ansatz 
is indeed correct at least up to $n=9$. However, given the very non-trivial fractions
that appear, it is more than natural to extend the validity of (\ref{fnlim}) to arbitrary $n>1$.

Exploiting this result, we then find that the large-$N$ behavior of the instanton part of the prepotential in the $\cN=2^*$ theory is
\begin{equation}
\cF^{\mathrm{inst}}\,\underset{N\to \infty} {\sim}~  N\,m^2\,\log \widehat \eta(q)   -Nm^2\sum_{n=2}^\infty \Big(\frac{m}{v}\Big)^{2n-2} \!\frac{\zeta(2n-2)}{n(2n+1)(2n+2)}\big(E_{2n-2}-1\big)~.
\label{Finsttot}
\end{equation}
Here all instanton sectors have been taken into account, and thus (\ref{Finsttot}) is exact in $q$. Differently from the pure
SU($N$) theory discussed in Section \ref{subsec:pure}, we see that in this case the dependence on $N$ is just in the overall prefactor. The instanton prepotential scales as $N$ but now, unlike in the pure $\cN=2$ case, the eigenvalues are distributed
in a region growing as $N$, while the masses of gauge bosons are kept finite in the limit. This instanton contribution is subleading 
with respect to the classical term that scales as $N^3$, but it gives the dominant contribution to  terms of order $m^4$ or 
higher in the prepotential.   

Formula (\ref{Finsttot}) is exact in $q$ and perturbative in $m$. Alternatively, this formula can be rewritten in a form where the exact 
$m$-dependence is displayed order by order in $q$. To this aim, 
one can use the expansion (\ref{eis2nq1}) of the Eisenstein series and perform the sum over $n$. After some easy algebra,
one finds the simple result
\begin{equation}
\cF^{\mathrm{inst}}\,\underset{N\to \infty} {\sim} ~\frac{N v^2}{2\pi^2}\sum_{k=1}^\infty \Big[
\cos\Big(\frac{2\pi k m}{v}\Big)-1\Big]\frac{1}{k^3}
\,\frac{q^{k}}{1-q^{k}}~.
\label{Finsttot3}
\end{equation}
Since all powers of the hypermultiplet mass have been resummed, this formula is exact in $m$. 
It is interesting to observe that  $\cF^{\mathrm{inst}}$ is periodic, i.e. $m \sim m+v$ and that it is always of order of $N$ except
for $m=v$ where all instanton contributions exactly cancel\footnote{This cancelation can be also seen directly from (\ref{zk}) noticing that poles in the integrand contains always the term $\chi_I=a_{i_0}$ for some $i$, which leads to a zero eigenvalue 
in the  numerator.}.  

We conclude this section by observing that when an $\Omega$-background is turned on, the remarkable cancellations of all terms
containing $E_2$ that led to (\ref{fnlim}), do not occur any longer. However, in the Nekrasov-Shatashvili limit $h\to 0$ 
with a non-vanishing $\epsilon$, we find
\begin{equation}
\begin{aligned}
f_2&  \,\underset{N\to \infty} {\sim} -N\, \frac{M^4}{v^2}\,\frac{\zeta(2)}{12}\,E_2~,\\
f_3 &  \,\underset{N\to \infty} {\sim} -N\, \frac{M^6}{v^4}\,\frac{\zeta(4)}{60}\Big(1-\frac{\epsilon^2}{2M^2}\Big)E_4~,\\
f_4 &  \,\underset{N\to \infty} {\sim} -N\,\frac{M^8}{v^6}\,\frac{\zeta(6)}{168}\Big(1-\frac{4}{3}\frac{\epsilon^2}{M^2}
+\frac{2}{3}\frac{\epsilon^4}{M^4}\Big)E_6
\end{aligned}
\label{fndef}
\end{equation}
where $M$ is defined in (\ref{M2}).
This seems to suggest that also in the Nekrasov-Shatshvili limit the same pattern of the undeformed theory occurs, 
with $f_n$ proportional to $E_{2n-2}$; however our current results are too limited to allow for a reliable extrapolation at this stage.

\section{Large-$N$ limit with the Wigner distribution}
\label{sec:Wigner}
In this section we study the large-$N$ behavior of the $\cN=2^*$ theory assuming that the eigenvalues
$a_i$ are real and distributed in an interval $[-\mu,+\mu]$ with the Wigner semi-circle law
\begin{equation}
\rho(x)= \frac{2}{\pi\mu^2}\,\sqrt{\mu^2-x^2}~,
\label{wigner}
\end{equation}
where the prefactor has been chosen in oder to have unit normalization:
\begin{equation}
\int_{-\mu}^{+\mu}\!dx\, \rho(x) = 1~.
\label{norm}
\end{equation}

 \subsection{The extremal distribution}

The Wigner distribution is interesting for several reasons. First, one can show that  it extremizes the exact prepotential of the $\cN=2^*$ gauge theory in the large-$N$ limit. To see this, we start from the exact formulas (\ref{f23}) for $\cF^{\mathrm{quant}}= 
\sum_{n=1}^\infty f_n$ plus the classical term (\ref {Fclass}) (with the Yang-Mills $\theta$-angle set to zero)
and vary it with respect to $a_i$. One gets
\begin{equation}
\begin{aligned}
\frac{\delta \cF }{\delta a_i} &=-\frac{8\pi^2}{g^2}\,a_i+M^2 C_1(i)+\frac{M^2(M^2+h^2)}{6}\, E_2 \, C_3(i)\\
&+
\frac{M^2(M^2+h^2)}{36}\Big[(2M^2+3h^2)E_2^2+\frac{1}{5}(2M^2-6\epsilon^2+3h^2)E_4\Big]
 \,C_5(i) \\
&~~~~-\frac{M^4(M^2+h^2)}{144}\big(E_2^2-E_4\big) \left[C_{23}(i)+C1_{320}(i) \right]+\ldots
\end{aligned}
\label{variation}
\end{equation}
where the sums $C_n(i)$, $C_{nm}(i)$, $C1_{mnp}(i)$ are defined in (\ref{sumsi}) and the dots stand for higher order terms
which have a similar structure.

In the large-$N$ limit when the eigenvalues $a_i$'s have a distribution $\rho$, sums over the discrete index $i$ 
are replaced by integrals over a continous variable $x$ according to
\begin{equation}
\label{sumint}
\sum_{i=1}^N f(a_i) ~\to~ N \,\ddashint_{-\mu}^{+\mu} \!dx\, \rho(x)\, f(x)~,
\end{equation}
where the symbol $\ddashint$ donotes that a suitable prescription has to be used in order to remove the divergent
contributions that arise when two eigenvalues get close leading to singularities along the integration path. As discussed in Appendix
\ref{subappa:Wigner}, we adopt the so-called Hadamard regularization (see for instance \cite{hadamard1} and references therein)
that is a generalization of the usual Cauchy Principal Value prescription particularly suited for our case, where singularities of
higher order appear in the various sums. Applying this prescription to the Wigner semi-circle law 
(\ref{wigner}), in Appendix \ref{subappa:Wigner} we show that
\begin{equation}
\label{Cniwigner}
C_n(i) ~\to~ \left\lbrace\begin{array}{ll}
+\frac{2N}{\mu^2}\,x&~~\mbox{for~$n=1$}~,\phantom{\Big |}\\
-\frac{2N}{\mu^2}&~~\mbox{for~$n=2$}~,\phantom{\Big |}\\
~~~\,0&~~\mbox{for~$n>2$}~.\phantom{\Big |}
\end{array}\right.
\end{equation}
Using this result and observing that, for instance $C_{nm}(i)=C_n(i)C_m(i)-C_{n+m}(i)$, we easily see that all terms containing sums   $C_{n_1 n_2 \ldots }(i), \ldots $ with at least one index $n_i\geq 3$  vanish. Thus, only the first two terms in (\ref{variation}) survive and lead to 
\begin{equation}
\frac{\delta \cF}{\delta a_i} ~\to~
-\frac{8\pi^2}{g^2}\,x+M^2 \frac{2N}{\mu^2}\,x
\label{variation2}
\end{equation}
which vanishes if
\begin{equation}
\mu^2= \frac{Ng^2}{4\pi^2}\,M^2~.
\label{muM}
\end{equation}
Thus, the Wigner semi-circle law (\ref{wigner}) with a mass scale $\mu$ given as in (\ref{muM}) is a stationary 
configuration of the full prepotential of the $\cN=2^*$ gauge theory in the large-$N$ limit.
This same result (but restricted to $\cF^{\mathrm{pert}}$) has been recently obtained in \cite{Buchel:2013id,Buchel:2013fpa} where the large-$N$ behavior of the  $\cN=2^*$ theory on $\mathbb{S}^4$ and its Wilson loops have been studied using localization techniques
\cite{Russo:2012ay,Russo:2013qaa,Russo:2013kea}. As suggested in \cite{Pestun:2007rz}, in this context
the sphere $\mathbb{S}^4$ can be mimicked by considering an $\Omega$ background whose 
parameters $\epsilon_1$ and $\epsilon_2$ are equal and related to the sphere radius $R$.
In our notations this means taking
\begin{equation}
\epsilon^2=4h^2=-\frac{4}{R^2}~.
\label{radius}
\end{equation}
Recalling the definition (\ref{M2}),  we can therefore rewrite (\ref{muM}) as
\begin{equation}
\mu^2= \frac{Ng^2}{4\pi^2 R^2}\,(m^2 R^2+1)
\label{muR}
\end{equation}
which exactly agrees with the findings of \cite{Buchel:2013id,Buchel:2013fpa}. This is also consistent with the analysis based
on the supergravity solution dual to the $\cN=2^*$ theory on $\mathbb{S}^4$ recently found in \cite{Bobev:2013cja} using holography.
Notice that by taking the limit $R\to\infty$, or equivalently $\epsilon,h\to0$, we recover the flat space description and can easily
reproduce the derivation of the Wigner distribution law originally presented in \cite{Buchel:2000cn} starting from the
so-called Pilch-Warner supergravity solution \cite{Pilch:2000ue}. We also observe that all these results imply that the interval 
$[-\mu,+\mu]$ in which the scalar eigenvalues are distributed has a width which grows as $\sqrt{N}$, differently from the
uniform distribution of the previous section whose width was growing linearly with $N$. 

\subsection{The large N limit}

We now want to evaluate the large-$N$ limit of the prepotential $\cF$ with the Wigner semi-circle law
(\ref{wigner}). To do so we need to estimate how the various sums behave at large $N$. As is clear from (\ref{sumint}),
a sum over a single index produces a factor of $N$; analogously multiple sums over $m$ indices become proportional to $N^m$.
Therefore, the large-$N$ limit is dominated by the terms with the largest number of indices summed over. {From} 
the explicit expressions of the prepotential coefficients $f_n$ presented in Section~\ref{secn:prep}, we easily realize that
those are the contributions containing the sums 
\begin{equation}
\label{C2'sbis}
C_{\underbrace{\mbox{\scriptsize{22\ldots2}}}_{\mbox{\scriptsize{$n-1$}}}}
\end{equation}
already encounterd in (\ref{C222n}). These sums involve $n$ summed indices and therefore are proportional%
\footnote{To check this, take into account that, according to their definitions (\ref{sums}), the quantities $C_{n_1n_2\ldots n_m}$ involve a sum over $m+1$ indices, while $C1_{n_1n_2\ldots n_m}$ , $C2_{n_1n_2\ldots n_m}$, $\ldots$ are sums over $m$ indices only.} to $N^n$.
Actually, using the Hadamard regularization prescription described in Appendix \ref{subappa:Wigner}, 
one can check that all sums except (\ref{C2'sbis}) vanish when evaluated with the Wigner distributions, 
so they can be discarded  in the large-$N$ limit.

Thus,  at large $N$ the $f_n$'s reduce to  
\begin{equation}
\begin{aligned}
f_1&= \frac{M^2}{4}\sum_{i\not=j} \log \left( \frac{a_{ij}}{\Lambda} \right)^2 +N\,(M^2+h^2)\log \widehat \eta(q)  ~, \\
f_2&=-\frac{M^2(M^2+h^2)}{24}\,E_2 \,C_2~,\\
f_3&=\ldots+\frac{M^4(M^2+h^2)}{576}\,\big(E_2^2-E_4\big) \,C_{22}~,\\
f_4&=\ldots-\frac{M^6(M^2+h^2)}{10368}\,\big(E_2^3-3E_2E_4+2E_6\big) \,C_{222}~,\\
f_5&=\ldots+\frac{M^8(M^2+h^2)}{165888}\,\big(E_2^4-6E_2^2E_4+8E_2E_6-3E_8\big) \,C_{2222}~,\\
f_6&=\ldots-\frac{M^{10}(M^2+h^2)}{2488320}\,\big(E_2^5-10E_2^3E_4+20E_2^2E_6-15E_2E_8+4E_{10}\big) \,C_{22222}~.
\end{aligned}
\label{fC2}
\end{equation}
Using the properties of the Eisenstein series collected in Appendix \ref{appb}, one can show that all the terms with $n>1$ 
can be nicely written as
\begin{equation}
f_{n}=\ldots-(-1)^n\frac{M^{2n-2}(M^2+h^2)}{24\,(n-1)!}\,D^{n-2}\!E_2~C_{\underbrace{\mbox{\scriptsize{22\ldots2}}}_{\mbox{\scriptsize{$n-1$}}}}
\label{fnC2}
\end{equation}
where $D=q\,d/dq$ is the logarithmic derivative with respect to the instanton weight. 
It is natural to think that such a formula is valid for any $n>1$. On the other hand, the large-$N$ behavior of the sums appearing above has been evaluated for the Wigner distribution in (\ref{C2nw}), 
which we rewrite here for convenience:
\begin{equation}
\label{C2nwf}
 C_{\underbrace{\mbox{\scriptsize{22\ldots2}}}_{\mbox{\scriptsize{$n-1$}}}}~
 \to ~ (-1)^{n-1}N^n\,\frac{2^{n-1}}{\mu^{2n-2}}~.
\end{equation}
Collecting (\ref{fC2}), (\ref{fnC2}) and (\ref{C2nwf}), the large-$N$ limit of the instanton prepotential 
can be written as
\begin{equation}
\begin{aligned}
\label{Fquantf}
\cF^{\mathrm{inst}} \,& \underset{N\to \infty} {\sim}   N\,(M^2+h^2)\Bigg[ \log \widehat \eta(q)   +  
\sum_{n=1}^\infty  
\,\frac{1}{24\,n!} \,\Big(\frac{2M^2N}{\mu^2}\Big)^{n}\,  D^{n-1}(E_2-1) \Bigg]~,
\end{aligned}
\end{equation}
We now use the definitions of the Eisenstein series given in Appendix \ref{appb} to write
\begin{equation}
\label{DE21}
D^{n-1} (E_2-1) = -24 \sum_{k=1}^\infty  k^{n-1}\sigma_1(k)\, q^{k}~;
\end{equation}
then, combining with (\ref{sigmaeta}), the sum over $n$ in (\ref{Fquantf})  can be performed leading 
to the remarkably simple formula
\begin{equation}
\label{Fquant3}
\cF^{\mathrm{inst}} \, \underset{N\to \infty} {\sim} \,  N\,(M^2+h^2) \log\widehat \eta(q_{\rm eff})    
\end{equation}
where
\begin{equation}
\label{shift}
 q_{\rm eff}=   \ee^{2 \pi \ii \tau+\frac{2 M^2 N }{\mu^2}  } \,=\,\ee^{-8\pi^2 /g^2+  \frac{2 M^2 N }{\mu^2}  + \ii \theta }~.
\end{equation}
Eq.~(\ref{Fquant3}) is exact in $q$, in the $\Omega$-background parameters $\epsilon,h$ and in the mass $m$. 
Interestingly, the instanton prepotential depends on $\mu$ only through the shift (\ref{shift}) in the gauge coupling 
constant! 
The finiteness of the effective gauge coupling $q_{\rm eff}$ requires that $\mu$ grows like $\sqrt{N} M$  in the large-$N$ limit. 
With this scaling, the instanton prepotential is of order $N$ and thus is subleading with respect to the classical and perturbative parts 
which, with the Wigner distribution, are $O(N^2)$. Like in the uniform distribution considered in the previous section, also in this
case there is no extra suppression in $N$ of the type that occurs for example in the pure SU$(N)$ theory (see (\ref{f1largeNfin})).
If we take the large-$N$ limit by keeping fixed the 't Hooft coupling, all instantons are exponentially suppressed at large $N$ and 
negligible, but if instead we keep fixed the effective Yang-Mills coupling, then the instanton sectors are which are weighted by  
powers of $q_{\rm eff}$, cannot be discarded.

\section{Conclusions}
\label{sec:concl}
We have generalized the modular anomaly equation of \cite{Billo:2013fi,Billo:2013jba} to $\cN=2^*$ theories with an arbitrary
number of colors in a generic $\Omega$-background, and used it to explicitly compute the coefficients of the prepotential
in a small-mass expansion, including all non-perturbative corrections. Quite remarkably, these coefficients
can be expressed as polynomials of quasi-modular functions of the gauge coupling, or better of the instanton weight $q$.
These expressions are $q$-exact and thus in principle 
could be used to analyze the behavior of the prepotential in the strong coupling regime where $q\to 1$. 

We have then studied the large-$N$ limit of these coefficients by taking $N\to\infty$
with the Yang-Mills coupling $g^2$ kept fixed. Differently from the usual 't~Hooft limit in which $q$ is exponentially suppressed
at large $N$, in our case $q$ remains finite so that instantons can not be discarded a priori. In this scenario it is therefore important 
to study the non-perturbative sector of the theory at large-$N$ and to take its effects into account.
In particular, we have investigated how the prepotential coefficients behave in this large-$N$ limit for two different 
distributions of scalar eigenvalues: a uniform distribution and the Wigner semi-circle law. 

In the case of the homogenous distribution, we have provided two alternative formulas for the large-$N$ prepotential: 
one, perturbative in the mass with coefficients given in terms of modular functions (see (\ref{Finsttot})), and
the other where the exact dependence on the mass is displayed order by order in $q$ (see (\ref{Finsttot3}).

In the case of the Wigner distribution, the large-$N$ limit of the prepotential leads to the remarkable simple formula (\ref{Fquant3}),
which is exact both in $q$ and in $m$, also in presence of an $\Omega$-background. It is interesting to 
notice that by taking the limit $M\to 0$ in this formula, one finds 
the $\cN=4$ instanton partition function \cite{Bruzzo:2002xf}, namely
\begin{equation}
Z^{\mathrm{inst}}=\ee^{-\frac{\cF^{\mathrm{inst}}}{h^2} }=\widehat\eta(q)^{-N}= \prod_{n=1}^\infty(1-q^n)^{-N}~,
 \end{equation}
in agreement with the general expectation that in the massless limit the $\cN=2^*$ theory becomes the $\cN=4$ super Yang-Mills theory.
Remarkably, the combined effects of turning on a mass and of selecting the Wigner distribution
amount to the simple replacements
\begin{equation}
 q \to q_{\rm eff}= \ee^{2 \pi \ii \tau+\frac{2 M^2 N }{\mu^2}  }    \quad ,\quad   h^2 \to h^2+M^2
\end{equation}
in the $\cN=4$ prepotential to obtain the $\cN=2^*$ expression.
 
We conclude by observing that our results can be applied also to the computation of the partition function for a gauge theory 
on $\mathbb{S}^4$. In fact, as it follows from (\ref{variation}) and (\ref{muM}), the Wigner distribution extremizes 
the prepotential also in presence of non-perturbative contributions and therefore it gives the leading contribution in a saddle-point  approximation of the prepotential at large $N$.
Moreover, the critical value $\mu^2=\frac{ g^2 N M^2  }{4\pi^2} $ for which the prepotential is extremized, is such that 
the  typical instanton suppression factor $\exp\left(-8\pi^2  /g^2\right)$ exactly balances an identical contribution coming from 
the $\mu$-dependent term, leaving a constant effective coupling $q_{\rm eff}$. Hence in this case instantons contribute 
with a $g$-independent term of order $M^2$, and
the relevant prepotential of the gauge theory on $\mathbb{S}^4$ at large $N$ comes entirely 
from the classical and 1-loop terms. These two contributions reduce to 
 \begin{equation}
 \begin{aligned}
 \cF&= -\frac{4\pi^2}{g^2}\sum_{i=1}^Na_i^2+  \frac{M^2}{2}\sum_{i\not=j} \log \left( \frac{a_{ij}}{\Lambda} \right) +\ldots
 \underset{N\to \infty} {\sim}  \,\frac{M^2 N^2}{4} \, \log  \left( \frac{g^2 N M^2}{\Lambda^2} \right)+\ldots
\end{aligned}
\end{equation}
where dots stand for subleading terms and all $g$-independent factors of order $M^2$ have been reabsorbed 
in the definition of $\Lambda$. Taking the third derivative of $\cF$ with respect to the mass $m$ and using (\ref{M2}), we obtain
\begin{equation}
\label{deriv}
\frac{d^3\cF}{dm^3}\,\underset{N\to \infty} {\sim}  \,N^2\,\frac{m\big(m^2-\ft{3}{4}\epsilon^2\big)}{\big(m^2-\ft{1}{4}\epsilon^2\big)^2}
\end{equation}
which, considering our conventions, is in perfect agreement with the results of \cite{Russo:2012ay,Russo:2013qaa,Russo:2013kea} obtained using localization, and of \cite{Bobev:2013cja} obtained from holography.

\vskip 1.5cm
\noindent {\large {\bf Acknowledgments}}
\vskip 0.2cm
We thank Yassen S. Stanev for collaboration in the early stage of this work and Igor Pesando for several useful discussions.
J.F.M. thanks the members of the University of Oxford and Imperial College London for their kind hospitality during the completion 
of this work.

The work of M.B., M.F. and A.L. is partially supported  by the Compagnia di San Paolo 
contract ``MAST: Modern Applications of String Theory'' TO-Call3-2012-0088.
The work of J.F.M. is partially supported by the ERC Advanced Grant n. 226455 ``Superfields'' and the Engineering and Physical Sciences Research Council, grant numbers  EP/I01893X/1 and EP/K034456/1.
The research of R.P. is partially supported by a Visiting Professor Fellowship from the University of Roma Tor Vergata, 
by the Volkswagen foundation of Germany, by a grant of the Armenian State Council of Science and
by the Armenian-Russian grant ``Common projects in Fundamental Scientific Research"-2013.

\vskip 1cm

\appendix
\section{The $\Gamma_2$-function}
\label{appd}
The Barnes double $\Gamma$-function is defined as
\begin{eqnarray}
\label{deflg2}
&&\log\Gamma_2(x|\epsilon_1,\epsilon_2)  = \frac{d}{ds}\left(  \frac{\Lambda^s}{\Gamma(s) } \int_0^\infty \frac{dt}{t}
\frac{t^s\, \ee^{-x t}}{(1-\ee^{-\epsilon_1 t} ) (1-\ee^{-\epsilon_2 t} )}\right)\Big|_{s=0}
\label{ge1e2}\\
&&~~~~~= \log \left( \frac{x}{\Lambda}\right)^2\Big(-\frac{1}{4} b_0 \, x^2  +\frac{1}{2}  b_1 \, x -\frac{b_2}{4}\Big)
+  \Big( \frac{3}{4} b_0 \, x^2 -b_1 \, x    \Big)  +\sum_{n=3}^\infty  \frac{b_n x^{2-n}}{n(n-1)(n-2)}
\nonumber
\end{eqnarray}
where the coefficients  $b_n$ are given by 
\begin{equation}
\label{cndef}
\frac{1}{(1-\ee^{-\epsilon_1 t} ) (1-\ee^{-\epsilon_2 t} ) }=\sum_{n=0}^\infty  \frac{b_n}{n!}\, t^{n-2}~.
\end{equation}
The first few of them are
\begin{equation}
\label{cnfirst}
b_0=\frac{1}{\epsilon_1 \epsilon_2}=\frac{1}{h^2}~,  \qquad   
b_1=\frac{\epsilon_1+\epsilon_2}{2 \epsilon_1 \epsilon_2}=\frac{\epsilon}{2h^2}~,\qquad b_2=\frac{\epsilon_1^2+3 \epsilon_1 \epsilon_2+\epsilon_2^2}{6\epsilon_1 \epsilon_2}=\frac{\epsilon^2+h^2}{6h^2}~.
\end{equation}

\section{Useful formulas for sums and their large-$N$ behavior}
\label{appa}
Here we give the definitions of the sum structures appearing in the prepotential coefficients $f_n$'s and study some of
their properties. We have
 \begin{equation}
 \begin{aligned}
 C_{n_1,n_2,\ldots, n_m} &=\sum_{ [i_1,i_2,\ldots, i_{m+1}] } \!\frac{1}{(a_{i_1{i_2}})^{n_1}}\frac{1}{(a_{i_1{i_3}})^{n_2}}\!\cdots\!
\frac{1}{(a_{i_1{i_{m+1}}})^{n_m}} \\
 C1_{n_1,n_2,\ldots, n_m} &=\sum_{ [i_1,i_2,\ldots, i_m] }\! \frac{1}{(a_{i_1 i_2})^{n_1}}\frac{1}{(a_{i_2 i_3})^{n_2}}\!\cdots\!
\frac{1}{(a_{i_{m} i_1})^{n_m}}\\
 C2_{n_1,n_2,\ldots, n_m} &=\sum_{ [i_1,i_2,\ldots, i_m] } \!\frac{1}{(a_{i_1 i_2})^{n_1}}\frac{1}{(a_{i_2 i_3})^{n_2}}
 \frac{1}{(a_{i_2 i_4})^{n_3}}\frac{1}{(a_{i_4 i_5})^{n_4}}\!\cdots\!\frac{1}{(a_{i_m i_1})^{n_m}}\\
C3_{n_1,n_2,\ldots, n_m} &=\sum_{ [i_1,i_2,\ldots, i_m] } \!\frac{1}{(a_{i_1 i_2})^{n_1}}\frac{1}{(a_{i_2 i_3})^{n_2}}
\frac{1}{(a_{i_2 i_4})^{n_3}}\frac{1}{(a_{i_2 i_5})^{n_4}}\frac{1}{(a_{i_5 i_6})^{n_5}}\!\cdots\!\frac{1}{(a_{i_m i_1})^{n_m}}
 \end{aligned}
 \label{sums}
 \end{equation}
where the symbol $[ i ,j, k,\ldots]$ denotes the sum over the positive integers with $i\neq j \neq k \cdots$. 
It is also convenient to consider the analogouus structures but without summing on one color index, namely
\begin{equation}
 \begin{aligned}
 C_{n_1,n_2,\ldots, n_m}(i)&=\sum_{ [j_1,j_2,\ldots, j_m]\not=i }
 \frac{1}{(a_{i{j_1}})^{n_1}}\frac{1}{(a_{i{j_2}})^{n_2}}\!\cdots\!\frac{1}{(a_{i{j_m}})^{n_m}} \\
 C1_{n_1,n_2,\ldots, n_m}(i) &=\sum_{ [j_1,j_2,\ldots, j_{m-1}]\not=i } 
 \frac{1}{(a_{i j_1})^{n_1}}\frac{1}{(a_{j_1 j_2})^{n_2}}\!\cdots\!
\frac{1}{(a_{j_{m-1} i})^{n_m}}
 \end{aligned}
 \label{sumsi}
 \end{equation}
and so and so forth. Actually, not all these sums are independent of each other, since there exist 
various algebraic identities among them which we are going to discuss.

\subsection{Identities}
\label{subappa:ident}
First of all, from the definition (\ref{sums}) it is straightforward to see that
\begin{equation}
C_n=0\quad\mbox{for~$n$~odd}~,
\label{Cnodd}
\end{equation}
and that
\begin{equation}
C_{11}=0
\label{C11}
\end{equation}
as a consequence of the relation
\begin{equation}
\frac{1}{a_{ij}a_{ik}}+\frac{1}{a_{ji}a_{jk}}+\frac{1}{a_{ki}a_{kj}}=0
\label{rel0}
\end{equation}
which holds for any $i$, $j$ and $k$. Multiplying (\ref{rel0}) by $1/(a_{ij})^2$ and then summing over
$[i,j,k]$, we easily find
\begin{equation}
2\,C_{31}-C1_{211}=0~;
\label{C31}
\end{equation}
likewise, multiplying (\ref{rel0}) by $1/(a_{ij}a_{ik})$ and then summing over $[i,j,k]$ we find
\begin{equation}
C_{22}+2\,C1_{211}=0~.
\label{C22}
\end{equation}
These two relations together imply
\begin{equation}
C_{22}+4\,C_{31}=0
\label{C22C31}
\end{equation}
which is used in the recursion relation to write $f_3$ as in (\ref{f23}).

This method can be easily generalized to derive many other identities.
For example, multiplying (\ref{rel0}) by $1/[(a_{ij})^{n-1}(a_{jk}^{m-1})(a_{ki}^{p-1})]$ and
summing over the indices, we get
\begin{equation}
C1_{n,m-1,p}+C1_{n,m,p-1}+C1_{n-1,m,p}=0~.
\label{C1}
\end{equation}
Setting $n=3$, $m=p=2$ we have
\begin{equation}
C1_{312}+C1_{321}+C1_{222}=0~;
\label{C1a}
\end{equation}
on the other hand setting $n=m=3$, $p=1$ and observing that $C1_{330}=-C_{33}$, we have
\begin{equation}
C1_{321}-C_{33}+C1_{231}=0~.
\label{C1b}
\end{equation}
Combining these last two relations and taking into account the cyclic properties $C1_{nmp}=C1_{mpn}=C1_{pnm}$, we obtain
\begin{equation}
C_{33}=-C1_{222}
\label{id0}
\end{equation}
which is the identity needed to write $f_4$ as in (\ref{f4}). Repeatedly using (\ref{C1}), one can check that
\begin{equation}
\begin{aligned}
C1_{npp}&=0\quad\mbox{for~$n$~odd}~,\\
C1_{422}&=2\,C_{62}+4\,C_{71}~,\\
C1_{622}&=2\,C_{82}+4\,C_{91}~,\\
C1_{442}&=C_{64}+2\,C_{73}+4\,C_{82}+8\,C_{91}~.
\end{aligned}
\label{id1}
\end{equation}
These relations are needed to cast $f_5$ and $f_6$ in the form presented in (\ref{f5}) and (\ref{f6}). 

Generalizing further these manipulations, we can prove a whole set of identities involving sums with more indices. Those which
are useful to check our explicit results are 
\begin{equation}
\begin{aligned}
&C1_{2222}-2\,C2_{2222}=-4\,C2_{3122}=2 \,C_{332}~,
\\
&C1_{2222}-8C2_{2222}=4C_{332}-4C_{62}+4\sum_{i=1}^N C_3(i)\,C1_{320}(i)~,
\\
&\frac{3}{2}C1_{4222}-C2_{2224}-\frac{1}{2} C2_{2422}-\frac{1}{2} C2_{4222}=C_{532}~,\\
&\frac{3}{2}C1_{4222}-\frac{1}{2}C2_{2224}-C2_{2422}+\frac{1}{2} C2_{4222}=-\frac{1}{2}C_{433}\phantom{\Big |}~,\\
&C1_{4222}-C2_{2422}+\frac{1}{2}C2_{4222}=\frac{1}{2}C_{442}~,\\
&\frac{5}{2}C1_{4222}-\frac{5}{2}C2_{2422}=\frac{1}{4}C1_{442}-C1_{622}+\sum_{i=1}^N C_3(i)\,C1_{322}(i)~.
\end{aligned}
\label{id2}
\end{equation} 

\subsection{Large-$N$ limit with a uniform distribution}
\label{subappa:unif}
We now  estimate the sums (\ref{sums}) in the large-$N$ limit with the distribution (\ref{vevsemiclassical}). The dependence 
on the scale $v$ is trivially fixed by dimensional arguments and so in the following we simply
use dimensionless quantities denoted with small $c$'s as opposed to capital $C$'s used for the sums (\ref{sums}). 
Thus we set
\begin{equation}
\label{defc}
c_{n_1,n_2,\ldots, n_m} \equiv v^{n_1+n_2 +\ldots n_m}\, C_{n_1,n_2,\ldots, n_m}~,
\end{equation}
and similarly for the other types of sums. 

To proceed, it is convenient to consider the quantities $C_n(i)$ introduced in (\ref{sumsi}), or better their dimensionless
counterparts which in this case become
\begin{equation}
c_n(i) \equiv v^n\, C_n(i) =  \sum_{j<i} \frac{1}{(i-j)^n} +  \sum_{j>i} \frac{1}{(i-j)^n}~.
\label{Cniu}
\end{equation}
Changing summation index, these quantities can be expressed in terms of (generalized) harmonic numbers:
\begin{equation}
\label{CCNiu2}
c_n(i) = 
\sum_{s=1}^{i-1}\frac{1}{s^n}+ (-1)^n \sum_{s=1}^{N-i}\frac{1}{s^n} = H(i-1,n) + (-1)^n H(N-i,n)~.
\end{equation}
In turn, the harmonic numbers can be expressed in terms of the Riemann and Hurwitz $\zeta$-functions as follows
\begin{equation}
\label{ghndef}
H(M,n)  = \sum_{s=1}^{M} \frac{1}{s^n} = \sum_{s=1}^\infty \frac{1}{s^n} - \sum_{s=M+1}^\infty \frac{1}{s^n}  
= \zeta(n) - \zeta(n,M+1) ~,
\end{equation}
so that 
\begin{equation}
\label{CNiu2}
c_n(i) = \big(1 +(-1)^n\big)\zeta(n) - \zeta(n,i) - (-1)^n \zeta(n,N-i+1)~.
\end{equation}
The asymptotics of the Hurwitz $\zeta$-function
\begin{equation}
\label{ash}
\zeta(n,M+1)\,\underset{M\to \infty} {\sim} \,
\frac{1}{(n-1)M^{n-1}} - \frac{1}{2 M^n} + \frac{n}{12 M^{n+1}} + \ldots
\end{equation}
implies that, for $n>1$, the $i$-dependent parts in (\ref{CNiu2}) are limited and rapidly approach zero for either $i$ growing 
toward $N$ or decreasing toward $1$. These  parts will not contribute when (products of) the $c_n(i)$'s are summed over $i$, 
thus greatly simplifying the evaluation of many sums. 

Let us give some details starting from $c_n$. By symmetry we have $c_n=0$ when $n$ is odd. 
For $n$ even, instead, using (\ref{CNiu2}) we find 
\begin{equation}
c_n = \sum_{i=1}^N c_n(i)=
2N\zeta(n)-2\sum_{i=1}^N\zeta(n,i)~.
\end{equation}
As argued above, the sum over the $i$-dependent parts is subleading in the large-$N$ limit; in fact one can easily 
show that   
\begin{equation}
\lim_{N\to \infty}\sum_{i=1}^N\zeta(n,i)=\zeta(n-1)~.
\end{equation}
We thus conclude that
\begin{equation}
c_n \underset{N\to \infty} {\sim}
\left\lbrace\begin{array}{ll}
2N\zeta(n)&~~\mbox{for~$n$~even}~,\\
0&~~\mbox{for~$n$~odd}~.
\end{array}\right.
\label{limcn}
\end{equation}

Let us now consider $c_{nm}$. If $n+m$ is odd, $C_{nm}$ vanishes when evaluated on the uniform distribution
(\ref{vevsemiclassical}). We can thus restrict to the case when $n+m$ is even. For any choice of the $a_i$'s, one can show the following
algebraic identity
\begin{equation}
 \label{CnCnm}
 \sum_{i=1}^N C_n(i)\,C_m(i)=  C_{nm}+C_{n+m}~,
 \end{equation}
from which we deduce that
\begin{equation}
\label{cnm}
c_{nm} = \sum_{i=1}^N c_n(i)\,c_m(i) - c_{n+m}~.
\end{equation}
If we substitute (\ref{CNiu2}) in this formula, we see that again all terms in the sum involving the $i$-dependent parts
give subleading contributions in the large-$N$ limit. Thus, using (\ref{limcn}) we obtain
\begin{equation}
c_{nm}\underset{N\to \infty} {\sim}
\left\lbrace\begin{array}{ll}
4N\zeta(n)\zeta(m)-2N\zeta(n+m)&~~\mbox{for~$n,m$~even}~,\\
-2N\zeta(n+m)&~~\mbox{for~$n,m$~odd}~.
\end{array}\right.
\label{limcnm}
\end{equation}
The only exception is when $n=m=1$ since in this case the subleading terms in $c_1(i)$ can give rise
to an $O(N)$ contribution in the sum. In fact one can show that
\begin{equation}
\sum_{i=1}^N\big(c_1(i)\big)^2=\sum_{i=1}^N\big(H(i-1,1)-H(N-i,1)\big)^2 \underset{N\to \infty} {\sim}2N\zeta(2)~;
\end{equation}
thus from (\ref{CnCnm}) and (\ref{limcn}) we obtain 
\begin{equation}
c_{11}(i) \underset{N\to \infty} {\sim} 0
\end{equation}
consistently with the fact that $C_{11}$ identically vanishes. 
As a further check we notice that from (\ref{limcnm}) we have
\begin{equation}
\begin{aligned}
&c_{31}\underset{N\to \infty} {\sim}- 2N\zeta(4)=-\frac{\pi^4}{45}N~,\\
&c_{22}\underset{N\to \infty} {\sim} 4N\zeta(2)^2-2N\zeta(4)=\frac{4\pi^4}{45}N~,
\end{aligned}
\end{equation}
so that $c_{22}+4c_{31}=0$, in perfect agreement with the identity (\ref{C22C31}) valid for all $N$'s.

It is clear that in a similar way one can deduce the limiting behaviour of all $c_{n_1n_2\ldots n_m}$'s. 
For instance, one can start from the identity
\begin{equation}
\label{CnCnmp}
 \sum_{i=1}^N C_n(i)C_m(i) C_p(i) =  C_{nmp} + C_{(n+m)p} + C_{(n+p)m} + C_{(p+m)n} + C_{n+m+p}
\end{equation}
and use it to express $c_{nmp}$ in terms of the already evaluated sums $c_n$ and $c_{nm}$, and of $\sum_i c_n(i) c_m(i) c_p(i)$, in which only the $i$-independent parts of (\ref{CNiu2}) contribute.
Other types of sums, like for instance $C1_{n_1n_2\ldots}$ or $C2_{n_1n_2\ldots}$, are related to the $C_{n_1n_2\ldots}$ by identities like those presented in Subsection \ref{subappa:ident}, and thus can also be evaluated in this way.

We now collect the large-$N$ behavior of all sums that appear in the prepotential coefficients used in the main text or given in
Appendix \ref{appc}, obtained via the procedure just described. We have
\begin{align}
c_n~\,&\underset{N\to \infty} {\sim} ~\,2N\zeta(n)~,\notag\\
c_{n_1n_2}~\,&\underset{N\to \infty} {\sim} ~\,N\,\big[4\zeta(n_1)\zeta(n_2)-2\zeta(n_1+n_2)\big]\phantom{\Bigg |}~,\notag\\
c_{n_1n_2n_3}~\,&\underset{N\to \infty} {\sim}~\, N\,\big[8\zeta(n_1)\zeta(n_2)\zeta(n_3)
-4\zeta(n_1+n_2)\zeta(n_3)-4\zeta(n_1+n_3)\zeta(n_2)\notag\\
&~~~~~~~~~~~~-4\zeta(n_2+n_3)\zeta(n_1)+2\zeta(n_1+n_2+n_3)\big]\phantom{\Big |}~,\label{sumcnN}\\
c_{n_1n_2n_3n_4}~\,&\underset{N\to \infty} {\sim} ~\,N\,\big[16\zeta(n_1)\zeta(n_2)\zeta(n_3)\zeta(n_4)
-8\zeta(n_1+n_2)\zeta(n_3)\zeta(n_4) \notag\\
&~~~~~~~~~~~~-8\zeta(n_1+n_3)\zeta(n_2)\zeta(n_4)-8\zeta(n_1+n_4)\zeta(n_2)\zeta(n_3)\phantom{\Big |}\notag \\
&~~~~~~~~~~~~-8\zeta(n_2+n_3)\zeta(n_1)\zeta(n_4)-8\zeta(n_2+n_4)\zeta(n_1)\zeta(n_3)\phantom{\Big |}\notag \\
&~~~~~~~~~~~~-8\zeta(n_3+n_4)\zeta(n_1)\zeta(n_2)+8\zeta(n_1+n_2+n_3)\zeta(n_4)\phantom{\Big |}\notag \\
&~~~~~~~~~~~~+8\zeta(n_1+n_2+n_3)\zeta(n_4)+8\zeta(n_1+n_2+n_3)\zeta(n_4)\phantom{\Big |}\notag \\
&~~~~~~~~~~~~+8\zeta(n_1+n_2+n_3)\zeta(n_4)+4\zeta(n_1+n_2)\zeta(n_3+n_4)\phantom{\Big |}\notag \\
&~~~~~~~~~~~~+4\zeta(n_1+n_3)\zeta(n_2+n_4)+4\zeta(n_1+n_4)\zeta(n_2+n_3)\phantom{\Big |}\notag \\
&~~~~~~~~~~~~-12\zeta(n_1+n_2+n_3+n_4)\big]\phantom{\Big |}\notag
\end{align}
where all the $n_i$'s are even. Ricursively using these relations, we can obtain the large-$N$ behavior of the sums with more indices.
In particular using the above rules, we can show that
\begin{equation}
c_{\underbrace{\mbox{\scriptsize{22\ldots2}}}_{\mbox{\scriptsize{$n$}}}}~\,\underset{N\to \infty} {\sim} ~\,N\alpha_n\,\zeta(2n)
\label{c222N}
\end{equation}
where the numerical coefficients $\alpha_n$ are given in the following table:
\begin{equation}
\label{alphatable}
\begin{tabular}{c|ccccccccc}
$n$ &1 & 2 &3 &4&5&6&7&8&9\\
\hline
$\phantom{\Big |}\alpha_n$&\,2~ & 8~&18~ &32~&48~&\normalsize{$\frac{43200}{691}$}&72~&$\frac{268800}{3617}$&
$\frac{3048192}{43867}$
\end{tabular}~~.
\end{equation}
Based on these values, and in view of the 1-instanton checks performed in Section \ref{subsec:n2*unif},
we may infer that these coefficient are actually given by
\begin{equation}
\label{alphares}
\alpha_n = (-1)^n \frac{(n-1)!}{(n+1)(2n+1)\zeta(1-2n)}~.
\end{equation}
Inserting this into (\ref{c222N}), we find
\begin{equation}
c_{\underbrace{\mbox{\scriptsize{22\ldots2}}}_{\mbox{\scriptsize{$n$}}}}~\,\underset{N\to \infty} {\sim} ~
(-1)^n \frac{N(n-1)!}{(n+1)(2n+1)}\,\frac{\zeta(2n)}{\zeta(1-2n)} =\frac{2N (2\pi)^{2n}n!}{(2n+2)!}
\label{c222Na}
\end{equation}
where the last equality follows from the relations
\begin{equation}
\zeta(2n)=(-1)^{n+1}\frac{(2\pi)^{2n}}{2(2n)!}\,B_{2n}~,\quad
\zeta(1-2n)=-\frac{1}{2n}\,B_{2n}
\label{zetabern}
\end{equation}
with $B_{2n}$ being the Bernoulli numbers.

Finally, using (\ref{sumcnN}) and  the identities (\ref{id0}), (\ref{id1}) and (\ref{id2})), we find
\begin{align}
c1_{222}\,&\underset{N\to \infty} {\sim} \,2 N\zeta(6)~,\quad
c1_{422}\,\underset{N\to \infty} {\sim} \,\frac{4}{3}N\zeta(8)~,\quad
c1_{622}\,\underset{N\to \infty} {\sim} \, \frac{6}{5}N\zeta(10)\phantom{\Big |}~,
\notag\\
c1_{442}\,&\underset{N\to \infty} {\sim} \,\frac{4}{5} N\zeta(10)~,\quad
c1_{2222}\,\underset{N\to \infty} {\sim} \, \frac{8}{3}N\zeta(8)~,\quad
c2_{2222}\,\underset{N\to \infty} {\sim} \,4N\zeta(8)\phantom{\Big |}~,\label{sumN} \\
c1_{4222}\,&\underset{N\to \infty} {\sim} \,\frac{8}{5} N\zeta(10)~,\quad
c2_{4222}\,\underset{N\to \infty} {\sim} \, \frac{12}{5}N\zeta(10)~,\quad
c2_{2422}\,\underset{N\to \infty} {\sim} \,2N\zeta(10)\phantom{\Big |}~,\notag \\
c2_{2224}\,&\underset{N\to \infty} {\sim} \,\frac{14}{5}N\zeta(10)~,\quad
c2_{22222}\,\underset{N\to \infty} {\sim} \,\frac{24}{5} N\zeta(10)~,\quad
c3_{22222}\,\underset{N\to \infty} {\sim} \,\frac{32}{5} N\zeta(10)\phantom{\Big |}~.\notag\\
\notag
\end{align}

\subsection{Large-$N$ limit with the Wigner distribution}
\label{subappa:Wigner}
We now estimate the behaviour of the sums (\ref{sums}) in the large-$N$ limit using 
the Wigner semi-circle distribution (\ref{wigner}) with a scale given by (\ref{muM}). 

Let us first consider the simple sums $C_n(i)$ defined in (\ref{sumsi}). It is easy to realize that they can all be obtained from the
following generating function
\begin{equation}
C(i)= \sum_{j\not=i}\frac{1}{a_i-a_j-\eta} = \sum_{n=1}^\infty C_n(i)\,\eta^{n-1}~.
\label{Ci}
\end{equation}
In the large-$N$ limit with the Wigner distribution this sum gets replaced according to
\begin{equation}
C(i) ~\to~ C(x) = 
N \dashint_{-\mu}^{+\mu}\!dy ~ \frac{\rho(y)}{x-y-\eta} = 
\frac{2N}{\pi \mu^2}\,\dashint_{-\mu}^{\mu}\!dy ~\frac{\sqrt{\mu^2-y^2}}{x-y-\eta}~,
\label{Cx}
\end{equation}
where the symbol $\dashint$ means that the integral has to be evaluated using the Cauchy Principal Value prescription if
$-\mu \leq x-\eta\leq \mu$. In the following, for simplicity we will often set $\mu=1$ since
the correct $\mu$-dependence can always be recovered using simple dimensional arguments. We then have
\begin{equation}
\dashint_{-1}^{+1}\!dy ~\frac{\sqrt{1-y^2}}{x-y-\eta} =  \lim_{\epsilon \to 0^+} \frac{1}{2} \Big[
 I_+(\epsilon) + I_-(\epsilon)\Big]
\label{cx1}
\end{equation}
where
\begin{equation}
 I_{\pm}(\epsilon)= \int_{-1}^{+1}\!dy~\frac{\sqrt{1-y^2}}{x-y-\eta \pm \ii \epsilon }
 = \ii \pi \sqrt{1-(x-\eta \pm \ii\epsilon)^2} +\pi (x-\eta \pm \ii\epsilon)~.
\label{ipm}
\end{equation}
These integrals have been computed using the techniques for integrals of multivalued functions and the residue theorem. 
Inserting this result into (\ref{cx1}) and reinstating the $\mu$-dependence, we get
\begin{equation}
C(x) = \frac{2N}{\mu^2}(x-\eta)~,
\label{cx2}
\end{equation}
which, after expanding in powers of $\eta$, yields
\begin{equation}
\label{Cnx}
C_n(x) = \left\lbrace\begin{array}{ll}
+\frac{2N}{\mu^2}\,x&~~\mbox{for~$n=1$}~,\phantom{\Big |}\\
-\frac{2N}{\mu^2}&~~\mbox{for~$n=2$}~,\phantom{\Big |}\\
~~~\,0&~~\mbox{for~$n>2$}~.\phantom{\Big |}
\end{array}\right.
\end{equation}
Notice that this is equivalent to define
\begin{equation}
\begin{aligned}
C_n(x) = \frac{(-1)^{n-1}}{(n-1)!}\,\frac{d^{n-1}}{dx^{n-1}}
\Bigg[N\,\dashint_{-\mu}^{+\mu}\!dy ~\frac{\rho(y)}{x-y}\Bigg]
~\equiv ~\frac{2N}{\pi \mu^{2}}
\,\ddashint_{-\mu}^{\mu}\!dy ~\frac{\sqrt{\mu^2-y^2}}{(x-y)^n}~,
\end{aligned}
\label{cnhad}
\end{equation}
where the symbol $\ddashint$ denotes the so-called Hadamard Finite Part prescription \cite{hadamard1}.
This prescription is a generalization of the Cauchy Principal Value which is suitable when the integrand exhibits high order singularities.
Essentially, the Hadamard Finite Part prescription amounts to minimally subtract all divergences from the integrand
and retain just the contributions coming from the residue at infinity. If the integrand has a simple pole on the integration path ($n=1$ in 
the above formula), the Hadamard prescription reduces to the Cauchy Principal Value, and if the integrand is regular it computes the
standard integral. For this reason from now on we always use the symbol $\ddashint$ even if sometimes it may be redundant.

Exploiting this result, the large-$N$ limit of the sums $C_n$ is then given by
\begin{equation}
C_n=\sum_{i=1}^N C_n(i) ~\to~N\,\ddashint_{-\mu}^{\mu}\!dx \,\rho(x)\,C_n(x)
= \left\lbrace\begin{array}{ll}
-\frac{2N^2}{\mu^2}&~~\mbox{for~$n=2$}~,\phantom{\Big |}\\
~~~\,0&~~\mbox{for~$n\not=2$}~.\phantom{\Big |}
\end{array}\right.
\end{equation}

Let us now consider the sums $C_{nm}$. Take for example the case $n=m=2$, corresponding to
\begin{equation}
\label{C2due}
C_{22} = \sum_{i\not=j_1\not=j_2}  \!\frac{1}{(a_{i}- a_{j_1})^{2}}\!\frac{1}{(a_{i}- a_{j_2})^{2}}~.
\end{equation}
In the large-$N$ limit, one integration variable (the one corresponding to the index $i$) appears in the integrand 
differently from those corresponding to $j_1$ and $j_2$ and thus the result may depend from the chosen order of integration. 
If, after setting $\mu=1$, we write
\begin{equation}
\label{C2duepo}
C_{22} ~\to~ N^3\,\ddashint_{-1}^{+1}\! dx\,\rho(x) ~
\ddashint_{-1}^{+1}\! dy_1\,\frac{\rho(y_1)}{(x-y_1)^2}~
\ddashint_{-1}^{+1}\! dy_2\,\frac{\rho(y_2)}{(x-y_2)^2}~,
\end{equation}
and use (\ref{cnhad}), we get
\begin{equation}
\label{C2duepores}
C_{22} ~\to~ 4N^3 ~.
\end{equation}
The same result is obtained with the choice
\begin{equation}
\label{C2duepop}
C_{22} ~\to~ N^3\,\ddashint_{-1}^{+1}\! dy_1\,\rho(y_1) ~
\ddashint_{-1}^{+1}\! dx\,\frac{\rho(x)}{(x-y_1)^2}~
\ddashint_{-1}^{+1}\! dy_2\, \frac{\rho(y_2)}{(x-y_2)^2} = 4N^3~.
\end{equation}
In both cases, nothing changes if we swap the order of integration over $y_1$ and $y_2$.
However, if instead we write 
\begin{equation}
\label{C2dueso}
C_{22} ~\to~ N^3\,\ddashint_{-1}^{+1}\! dy_1\,\rho(y_1) ~
\ddashint_{-1}^{+1}\! dy_2\,\rho(y_2)~
\ddashint_{-1}^{+1}\! dx\,\frac{\rho(x)}{(x-y_1)^2(x-y_2)^2}~,
\end{equation}
we immediately get
\begin{equation}
\label{C2duesores}
C_{22}\to 0
\end{equation}
since the Hadamard Finite Part of the integral over $x$ vanishes. Again, we get the same result
if we invert the order of integration on $y_1$ and $y_2$.

To overcome this problem and treat all integration variables on the same footing as one does with the discrete indices 
in the original finite-$N$ sum (\ref{C2due}), we have to symmetryze over all possible orderings of integration.
Introducing a simplified notation where
\begin{equation}
\label{notint}
\ddashint_{x_1x_2\ldots x_n}~~ \quad\mbox{stands for}~~\qquad
\ddashint_{-1}^1\! dx_1\,\rho(x_1) ~\,\ddashint_{-1}^1\! dx_2\rho(x_2) \ldots~\ddashint_{-1}^1\! dx_n\rho(x_n) ~, 
\end{equation}  
we denote the symnmetrized integration as
\begin{equation}
\label{symint}
\ddashint_{\{x_1x_2\ldots x_n\}} \equiv ~~\cN_n \sum_{P\in S_n}~ \ddashint_{x_{P(1)} x_{P(2)} \ldots 
x_{P(n)}}
\end{equation}
where $\cN_n$ is a suitable normalization coefficient to be fixed and $P$ is a permutation.

Evaluating $C_{22}$ with this symmetrized prescription and reinstating the $\mu$-dependence, we get
\begin{equation}
\label{C22sym}
C_{22} ~\to~~ \ddashint_{\{x\,y_1y_2\}}\frac{1}{(x-y_1)^2(x-y_2)^2}\,= \, \cN_3\,\frac{16 N^3}{\mu^4}~.
\end{equation}
Indeed, as we have seen before, there are four different integration orderings each of which yields a contribution proportional $4N^3$ 
and two orderings which do not contribute. This same procedure can be straightforwardly applied to other sums; for example
we find
\begin{equation}
\label{C31sym}
C_{31} ~\to~~ \ddashint_{\{x\,y_1y_2\}}\frac{1}{(x-y_1)^3(x-y_2)}\,= \, -\cN_3\,\frac{4 N^3}{\mu^4}~.
\end{equation}
In this case, out of the six different integration orderings, only one of them gives a non-vanishing contribution proportional to $-4N^3$.
Notice that the results (\ref{C22sym}) and (\ref{C31sym}) are perfectly consistent with the identity $4C_{31}+C_{22}=0$, 
which is valid for any $N$ (see (\ref{C22C31})). Likewise, one can check that
\begin{equation}
\label{C11sym}
C_{11} ~\to~~ \ddashint_{\{x\,y_1y_2\}}\frac{1}{(x-y_1)(x-y_2)}\,= \, 0
\end{equation}
as it should be in view of (\ref{C11}). It worth stressing that in this case all six different orderings give non-vanishing contributions to the integral, and it is only after summing them all that we obtain the correct result.
We can apply this prescription to evaluate the large-$N$ behavior of all types of sums defined in (\ref{sums}). In all cases one can
verify that the results are always consistent with the relations derived in Appendix \ref{subappa:ident}; these non-trivial checks put the
entire procedure on a very solid basis.

To fix the normalization we can exploit the algebraic identity
\begin{equation}
C_{22}=\sum_{i=1}^N C_2(i)\,C_2(i)-\sum_{i=1}^N C_4(i)~,
\label{C22id}
\end{equation}
which in the large-$N$ limit becomes
\begin{equation}
C_{22}~\to~ N\,\ddashint_{-\mu}^{+\mu}\!dx~ \rho(x)\,\big[C_2(x)\,C_2(x)-C_4(x)\big] =\frac{4N^3}{\mu^4}
\end{equation}
where in the last step we used (\ref{cnhad}). Comparing with (\ref{C22sym}), we can fix the normalization factor: $\cN_3=1/4$.

The generalization of this procedure to other sums is straightforward. In particular, for the sums (\ref{C2'sbis}), that are relevant
to study the large-$N$ limit of the $\cN=2^*$ prepotential,  we find 
\begin{equation}
\label{C2nw}
C_{\underbrace{\mbox{\scriptsize{22\ldots2}}}_{\mbox{\scriptsize{$n-1$}}}}~
\to ~ (-1)^{n-1}N^n\,\frac{2^{n-1}}{\mu^{2n-2}}~.
\end{equation}

\section{Eisenstein series and their modular properties}
\label{appb}
The Eisenstain series $E_{2n}$ are holomoprhic functions of $\tau$ defined as
\begin{equation}
\label{defeis}
E_{2n} = \frac{1}{2\zeta(2n)}\sum_{m,n\in\mathbb{Z}^2\setminus \{0,0\}} \frac{1}{(m+n\tau)^{2n}}~.
\end{equation}
For $n>1$, they are \emph{modular forms} of degree $2n$: under an $\mathrm{SL}(2,\mathbb{Z})$ transformation
\begin{equation}
\label{sl2z}
\tau \to \tau^\prime = \frac{a\tau + b}{c\tau +d}~~~\mbox{with}~~
a,b,c,d \in\mathbb{Z}~~~\mbox{and}~~
ad-bc=1
\end{equation}
one has
\begin{equation}
\label{eissl2z}
E_{2n}(\tau^\prime) = (c\tau + d)^{2n} E_{2n}(\tau)~.
\end{equation}
For $n=1$, the $E_2$ series is instead \emph{quasi-modular}:
\begin{equation}
\label{eis2sl2z}
E_{2}(\tau^\prime) = (c\tau + d)^{2} E_{2}(\tau) + \frac{6}{\ii\pi}c(c\tau + d)~.
\end{equation}

All the \emph{modular} forms of degree $2n>6$ can be expressed in terms of $E_4$ and $E_6$; in particular, this is true of
the Eisenstein series:
\begin{equation}
\label{e2ne4e6}
\begin{aligned}
E_8 & = E_4^2~,\\
E_{10} & = E_4 E_6~,\\
691 E_{12} & = 441 E_4^3 + 250 E_6^2~,\\
\ldots & \phantom{=} \ldots~.
\end{aligned}
\end{equation}  
\emph{Quasi-modular} forms of higher degree can be expressed as polynomials in $E_2$, $E_4$ and $E_6$.

The Eisenstein series admit a Fourier expansion which, in terms of $q=\exp(2\pi\ii\tau)$, takes the form 
\begin{equation}
E_{2n}=1+\frac{2}{\zeta(1-2n)}\sum_{k=1}^\infty \sigma_{2n-1}(k) q^{k}~,
\label{eis2nq0}
\end{equation}
where $\sigma_p(k)$ is the sum of the $p-th$ powers of the divisors of $k$. 
In particular, this amounts to
\begin{equation}
\label{e246q}
\begin{aligned}
E_2 & = 1 - 24 \sum_{k=1}^\infty \sigma_1(k) q^{2k} = 1 - 24 q - 72 q^2 - 96 q^3 + \ldots~,\\
E_4 & = 1 + 240 \sum_{k=1}^\infty \sigma_3(k) q^{2k} = 1 + 240 q + 2160 q^2 + 6720 q^3 + \ldots~,\\
E_6 & = 1 - 504 \sum_{k=1}^\infty \sigma_5(k) q^{2k} =1 - 504 q - 16632 q^2 - 122976 q^3 + \ldots~.
\end{aligned}
\end{equation}
The Fourier expansion \eq{eis2nq0} can also be rewritten as
\begin{equation}
E_{2n}
=1+\frac{2}{\zeta(1-2n)}\sum_{k=1}^\infty k^{2n-1}\,\frac{q^{k}}{1-q^{k}}
= 1+\frac{(-1)^n(2\pi)^{2n}}{(2n-1)!\,\zeta(2n)}\sum_{k=1}^\infty k^{2n-1}\,\frac{q^{k}}{1-q^{k}}~,
\label{eis2nq1}
\end{equation}
where in the second step we  used the relations (\ref{zetabern}).

The $E_2$ series is related to Dedekind's eta function
\begin{equation}
\label{etadef}
\eta(q) = q^{1/24}\prod_{k=1}^\infty (1 - q^{k}).
\end{equation}
Indeed, taking the logarithm of this definition we get
\begin{equation}
\label{logeta}
\log\left(\frac{\eta(q)}{q^{1/24}}\right) 
\equiv \log\widehat\eta(q)
= \sum_{r=1}^\infty \log\left(1 - q^{r}\right)
= - \sum_{k=1}^\infty \frac{\sigma_1(k)}{k} q^{k}~.
\end{equation}
If we apply now to this relation the derivative  operator $D = q d/dq$ we get
\begin{equation}
\label{Dlogeta}
D \log\widehat\eta = - \sum_{k=1}^\infty \sigma_1(k) q^{k} =
\frac{E_2 - 1}{24}~. 
\end{equation}
Applying repeatedly the operator $D$ to this last expression we also find
\begin{equation}
\label{repDe2}
D^{n-1} (E_2 -1) = -24 \sum_{k=1}^\infty k^{n-1} \sigma_1(k) q^{k}~.
\end{equation}
Finally, we also have
\begin{equation}
\label{De2e4d6}
\begin{aligned}
DE_2 & = \frac{1}{12} \left(E_2^2 - E_4\right)~,\\
DE_4 & = \frac{1}{3}  \left(E_2 E_4 - E_6\right)~,\\
DE_6 & = \frac{1}{2}  \left(E_2 E_6 - E_4^2\right)~.
\end{aligned}
\end{equation}

\section{Explicit expressions for $f_4$, $f_5$ and $f_6$}
\label{appc}
Exploiting the recursion relation (\ref{recursion}) and following the procedure outlined in the main text, we have computed
the coefficients $f_4$, $f_5$ and $f_6$ of the prepotential. 
Their explicit expressions are
\begin{align}
f_4=&-\frac{M^2(M^2+h^2)}{2592}\Big[\big(5M^4+17 M^2 h^2+15 h^4\big)E_2^3\notag\\
&~~~~~+\frac{3}{80}\big(64 M^4+256 M^2 h^2+240 h^4-192 M^2\epsilon^2-480\epsilon^2h^2+15\epsilon^4\big)E_2 E_4\notag\\
&~~~~~+\frac{1}{35}\big(11M^4+59 M^2 h^2+60 h^4-108 M^2\epsilon^2-270\epsilon^2h^2+45 \epsilon^4\big) E_6
\Big] C_6\notag\\
&+\frac{M^4(M^2+h^2)}{1728}\Big[\big(2M^2+3h^2\big)E_2^3-\frac{3}{5}\big(2M^2+4 \epsilon^2+3h^2\big)E_2 E_4\notag\\
&~~~~~-\frac{2}{5}\big(2M^2-6\epsilon^2+3h^2\big)E_6\Big] C_{42} \notag\\
&-\frac{M^4(M^2+h^2)}{10368}\Big[\big(M^2+4h^2\big)E_2^3-\frac{3}{5}\big(M^2+12 \epsilon^2+4h^2\big)E_2 E_4\notag\\
&~~~~~-\frac{2}{5}\big(M^2-18\epsilon^2+4h^2\big)E_6\Big] C1_{222} \notag\\
&-\frac{M^6(M^2+h^2)}{10368}\,\big(E_2^3-3E_2E_4+2E_6\big) \,C_{222}~,
\label{f4}
\end{align}
and, in the limit $\epsilon,h\to 0$,
\begin{align}
f_5=&-\frac{m^{10}}{10368}\Big(7 E_2^4+\frac{28}{5}E_2^2 E_4+\frac{44}{35}
E_2E_6+\frac{19}{35}E_8\Big)\,C_8\notag\\
&+\frac{m^{10}}{10368}\Big(5 E_2^4-E_2^2 E_4-\frac{18}{7}
E_2E_6-\frac{10}{7}E_8\Big)\,\Big(C_{62}+\frac{13}{30}C1_{422}\Big)\notag\\
&-\frac{m^{10}}{6912}\big( E_2^4-2E_2^2 E_4+E_8\big)\,\Big(C_{422}+\frac{7}{36}C1_{422}+\frac{5}{24}C1_{2222}
-\frac{1}{6}C2_{2222}\Big)\notag\\
&+\frac{m^{10}}{165888}\,\big(E_2^4-6E_2^2E_4+8E_2E_6-3E_8\big) \,C_{2222}~,
\label{f5}
\end{align}
and
\begin{align}
f_6=&-\frac{m^{12}}{25920}\Big(7 E_2^5+8 E_2^3E_4+\frac{33}{14}E_2^2 E_6+\frac{68}{35}
E_2E_8+\frac{37}{110}E_{10}\Big)\,C_{10}\notag\\
&+\frac{m^{12}}{31104}\Big(7 E_2^5+\frac{7}{5} E_2^3E_4-\frac{17}{5}E_2^2 E_6-\frac{18}{5}
E_2E_8-\frac{7}{5}E_{10}\Big)\notag\\
&~~~~\Big(C_{82}+\frac{17}{56}C1_{622}+\frac{71}{112}C1_{442}\Big)\notag\\
&-\frac{m^{12}}{62208}\Big(5 E_2^5-\frac{13}{2} E_2^3E_4-\frac{7}{2}E_2^2 E_6+\frac{3}{2}
E_2E_8+\frac{7}{2}E_{10}\Big)\notag\\
&~~~~\Big(C_{622}+\frac{11}{30}C1_{622}+\frac{43}{180}C1_{442}+\frac{109}{45}C1_{4222}
+\frac{64}{45}C2_{4222}-\frac{24}{15}C2_{2422}-\frac{5}{9}C2_{2224}\Big)\notag\\
&+\frac{m^{12}}{62208}\big(E_2^5-4 E_2^3E_4+2E_2^2 E_6+3
E_2E_8-2E_{10}\big)\notag\\
&~~~~\Big(C_{4222}+\frac{1}{12}C1_{622}-\frac{13}{72}C1_{442}+\frac{11}{18}C1_{4222}
-\frac{1}{4}C1_{22222}\notag\\
&~~~~~~~~~~~~~~+\frac{31}{36}C2_{4222}-\frac{1}{2}C2_{2422}-\frac{5}{18}C2_{2224}+
\frac{5}{8}C2_{22222}-\frac{1}{4}C3_{22222}\Big)\notag\\
&-\frac{m^{12}}{2488320}\,\big(E_2^5-10E_2^3E_4+20E_2^2E_6-15E_2E_8+4E_{10}\big) \,C_{22222}~.
\label{f6}
\end{align}

\providecommand{\href}[2]{#2}\begingroup\raggedright\endgroup

\end{document}